\documentclass[review,12pt]{elsarticle}
\journal{IJSS}
\usepackage{graphicx}
\usepackage[letterpaper]{geometry}
\usepackage{subfig}	
\usepackage{amssymb,amsfonts}
\usepackage{mathtools}
\usepackage{xfrac}
\usepackage{hyperref}
\usepackage{natbib}
\graphicspath{{./Figures/}} 
\usepackage{amsmath,amssymb,amsthm}
\usepackage{bm}
\usepackage{color}
\usepackage[normalem]{ulem}

\newcommand{\bit}{\begin{itemize}}
\newcommand{\eit}{\end{itemize}}
\newcommand{\benu}{\begin{enumerate}}
\newcommand{\eenu}{\end{enumerate}}
\newcommand{\be}{\begin{equation}}
\newcommand{\ee}{\end{equation}}
\newcommand{\bea}{\begin{eqnarray}}
\newcommand{\eea}{\end{eqnarray}}
\newcommand{\bean}{\begin{eqnarray*}}
\newcommand{\eean}{\end{eqnarray*}}
\newcommand{\ben}{\begin{equation*}}		
\newcommand{\een}{\end{equation*}}

\newlength{\figwidth}
\newlength{\figwidthtwo}
\newlength{\figwidththree}
\newcommand{\fref}[1]{Fig.\,\ref{#1}}
\newcommand{\tref}[1]{Table\,\ref{#1}}
\newcommand{\eref}[1]{Eq.\,(\ref{#1})}

\newcommand{\sref}[1]{Sec.\!~\ref{#1}}

\newcommand{\cref}[1]{Ref.\,\cite{#1}}

\newcommand{\vs}{{\it vs.}\! }

\newcommand{\eg}{{\it e.g.}\! }

\newcommand{\etal}{{\it et al.}\! }
\newcommand{\apriori}{{\it a priori} }

\newcommand{\Cbb}{\mathbb{C}}

\newcommand{\epsilonb}{\boldsymbol{\epsilon}}
\newcommand{\sigmab}{\boldsymbol{\sigma}}

\renewcommand{\sb}{\mathbf{s}}

\newcommand{\Ib}{\mathbf{I}}

\newcommand{\tr}{\operatorname{tr}}

\newcommand{\dev}{\operatorname{dev}}

\DeclareMathOperator*{\argmax}{arg\,max}

\begin{document}
\begin{frontmatter}
\title{\bf Sensitivity of void mediated failure to geometric design features of porous metals}

\author{G.H. Teichert}
\address{{\it University of Michigan, Ann Arbor, MI 48109}}
\author{M. Khalil}
\address{{\it Sandia National Laboratories, Livermore, CA 94551}}
\author{C. Alleman}
\address{{\it Sandia National Laboratories, Livermore, CA 94551}}
\author{K. Garikipati}
\address{{\it University of Michigan, Ann Arbor, MI 48109}}
\author{R.E. Jones\corref{COR}}
\cortext[COR]{corresponding: rjones@sandia.gov}
\address{{\it Sandia National Laboratories, Livermore, CA 94551}}

\begin{abstract}
Material produced by current metal additive manufacturing processes is susceptible to variable performance due to imprecise control of internal porosity, surface roughness, and conformity to designed geometry.
Using a double U-notched specimen, we investigate the interplay of nominal geometry and porosity in determining ductile failure characteristics during monotonic tensile loading.
We simulate the effects of distributed porosity on plasticity and damage using a statistical model based on populations of pores visible in computed tomography scans and additional sub-threshold voids required to match experimental observations of deformation and failure.
We interpret the simulation results from a physical viewpoint and provide statistical models of the probability of failure near stress concentrations.
We provide guidance for designs where material defects could cause unexpected failures depending on the relative importance of these defects with respect to features of the nominal geometry.

\end{abstract}

\end{frontmatter}

\section{Introduction} \label{sec:introduction}

Metal additive manufacturing (AM) is a relatively recent and now widespread manufacturing process that enables new levels of geometric complexity in designs.
In its current state of development, issues with uncontrolled porosity and surface roughness \cite{heckman2020automated} prevent AM from being adopted for certain high strength and high consequence applications.
Furthermore, depending on the scale of a part relative to the particles comprising the granular feedstock, adherence to the designed geometry can be poor \cite{garino2004mechanical,roach2020size}.
For example, surface roughness can be comparable to small, high curvature geometric features that result from sophisticated topology optimization and more mundane corners and fillets \cite{brackett2011topology,deng2017concurrent,vanderesse2018measurement,solberg2019fatigue,deng2017concurrent,delrio2020shoulder}, preventing the precise manufacture of some designs.  

In our previous work \cite{khalil2019modeling} we developed a model of porous metal deformation and failure tuned to experimental data from AM 17-4{H} steel \cite{salzbrenner2017high}.
That work employed Bayesian techniques to represent the observed variable mechanical response due to porosity and other microstructural effects.
In this effort we utilize the previously calibrated model together with the process we developed to generate porosity realizations to examine how stress concentrations due to geometric design features interact with material porosity.
To establish quantitative relationships between geometric and material effects, we focus on a double U-notched specimen, a test geometry commonly used to explore the influence of designed geometric features and inherent defects on ductile failure.
With a porous material there is a competition among the stress concentrations induced by notches and other geometric features and those due to the distributed porosity, which adds complexity to the design process.
In this work we quantify the probability of unexpected failure using statistical models based on a large sample of the porosity realizations.
We quantify the probability of failure as a function of the notch geometry, associating certain features of the failure distribution with the stress-inducing notches and others with the background density of pores. 

Although we do not apply reliability methods to the data, the work falls in the general context of design under uncertainty.
There is a large body of literature on the probabilistic failure of brittle materials \cite{trustrum1979estimating,trustrum1983applicability,bazant1991statistical,fok2001numerical,becker1988void}, in large part focused on the applicability of the Weibull and related distributions.
This is a mature field of study that has recently been applied to advanced materials, such as silicon based micro-electro-mechanical systems (MEMS) \cite{jadaan2003probabilistic,mccarty2007description,cook2019predicting,delrio2020shoulder}.
There has also been focus on the uncertainty due to the ductility and porosity of metals, AM or otherwise.
In fact, the uncertainty of the mechanical response of porous metals such as cast iron has long been been a topic of interest \cite{rice1993comparison,becker1988void,hardin2013effect}.
Notably, Yin \etal \cite{yin2009efficient} presented an integrated framework to perform design under material uncertainty.
They propagated the mean and standard deviation of damage based on a random field of initial damage and demonstrated a reliability-based method with the design of a cast control arm. 
Gurson-type constitutive models that they, and we, have employed have shown success in representing the complex failure process \cite{becker1988void,allison2013plasticity,shakoor2018ductile}.
For instance, Becker \etal \cite{becker1988void} compared predictions of a micro-mechanically based model to tensile data for steel  U-notched cylinders.
The model predictions up to and including failure compared favorably to the experimental data.
Allison \etal \cite{allison2013plasticity} collected comprehensive experimental data on the effects of porosity on the plasticity and damage of FC-0205 steel alloy made with powder metallurgy.
The associated modelling effort focused on the complex behavior of nearby void coalescence leading to failure.
Shakoor \etal \cite{shakoor2018ductile} studied the influence of particle debonding and fragmentation on void coalescence. 
Some other studies have focused on the statistical aspects of failure of porous metals.
Boyce \etal \cite{boyce2017extreme} were able to perform tension tests of 1000 nominally identical specimens of AM material.
With this large statistical sample they were able to capture rare failures associated with initial pore forming networks making the sample more susceptible to failure.
Solberg \etal \cite{solberg2019fatigue} investigated the fatigue behavior of AM 317 stainless steel with tension experiments.
They discovered a transition from surface initiated fatigue to internally initiated fatigue.
They were able to relate the trends in failure locations and behavior to the geometry of the specimen notch and the resulting stress concentration factors.
In a related effort, we use thousands of realizations of porous AM 17-4 PH specimens to explore the interplay of geometric design features and explicitly modeled pores in this work.
With this extensive exploration, which is necessary to sample rare failure events, we also find a transition in failure behavior and location with geometry and stress concentration.
We utilize these observations in a statistical model that characterizes the contributions of the stress concentration versus the background of randomly located pores.
This model enables estimation of the likelihood of failure, either as expected near a stress concentration versus randomly due to a critical pore not near the stress concentration.
This information has significant value in design.
We find that a significant likelihood of unexpected failures only occurs in a limited and bounded regime of mild stress concentrations.

In \sref{sec:simulation} we review the computational model we developed and calibrated in \cref{khalil2019modeling}.
In \sref{sec:results} we describe the results of parametric study of notches of various widths and depths and examine failure mechanisms.
Then in \sref{sec:model} we develop a statistical model of the failures observed across a range of notch geometries.
Finally, in \sref{sec:conclusion}, we summarize the findings and design guidelines.

\section{Simulation Method} \label{sec:simulation}

As discussed in the Introduction, the performance of AM materials can be significantly degraded by uncontrolled porosity.
This porosity arises from a variety of physical phenomena active in the AM process, giving rise to distributions of defects of varying size and character.
Thus it is difficult to represent the effects of this porosity with a bulk quantity such as overall material density.
As in our previous work \cite{khalil2019modeling}, we employ a model of distributed porosity which accurately reproduces the sizes and spatial correlations of defects observed in the material.
This two-level model represents pores above a given size threshold explicitly in the finite element meshes, and sub-threshold pores are modeled with an initial damage field.
The threshold is determined by the size of pores visible with current computed tomography (CT) resolution.
In this model sub-threshold porosity evolves as damage; and, highly damaged elements act like voids and therefore add to the influence of the explicit porosity.
In the following we will refer to explicitly represented pores as {\it pores} and the porosity implicit in the constitutive model as {\it voids}.

\subsection{Constitutive Model} \label{sec:material_model}
In the previous study \cite{khalil2019modeling} we calibrated an isothermal variant of the Bammann-Chiesa-Johnson (BCJ) viscoplastic damage model \cite{bammann1996modeling,horstemeyer1999void,brown2012validation,karlson2016sandia} to experimental data on AM 17-4 PH stainless steel tension specimens loaded to failure.
We employ that calibrated model to represent the deformation and failure behavior of the specimens in this study.
In the BCJ constitutive model, stress $\sigmab$ follows a linear elastic rule:
\begin{equation} \label{eq:stress}
\sigmab = (1-\phi) \Cbb ( \epsilonb - \epsilonb_p ) \ ,
\end{equation}
where $\Cbb$ is the isotropic elastic modulus tensor with components 
\begin{equation}
[ \Cbb ]_{ijkl} = \frac{E}{(1+v)} \left( \frac{v}{(1-2v)} \delta_{ij}\delta_{kl} + \frac{1}{2} (\delta_{ik}\delta_{jl} + \delta_{il}\delta_{jk}) \right)
\end{equation}
which depend on Young's modulus $E$ and Poisson's ratio $v$.
Here $\epsilonb$ is the total strain, $\epsilonb_p$ is the plastic strain, and $\epsilonb - \epsilonb_p$ is defined as the elastic strain.
The void fraction is denoted by $\phi$, and its evolution effects a reduction of stiffness of the material via \eref{eq:stress}.

The plastic strain ${\epsilonb}_p$ evolves according to the viscoplastic flow rule:
\begin{equation} \label{eq:dotep}
\dot{\epsilonb}_p = \sqrt{\frac{3}{2}} f \sinh^n \left( \frac{\sigma/(1-\phi) - \kappa}{Y} - 1 \right) \frac{\sb}{\| \sb \|} \ .
\end{equation}
The evolution of the isotropic hardening variable $\kappa$ is governed by:
\begin{equation} \label{eq:hardening}
\dot{\kappa} = (H-R \kappa) \, \dot{\epsilon}_p
\end{equation}
with $\kappa(t=0) = \kappa_0$.
Here, $\sb = \dev \sigmab$ is the deviatoric stress, $\sigma_\text{vm} \equiv \sqrt{\frac{3}{2} \sb\cdot\sb}$ is the von Mises stress, and $\dot{\epsilon}_p = \sqrt{\frac{2}{3} \dot{\epsilonb}_p \cdot \dot{\epsilonb}_p}$ is the equivalent plastic strain rate.
The model parameters are the yield stress $Y$, the flow coefficient $f$, the flow exponent $n$,  the hardening modulus $H$, and the recovery coefficient $R$.

The damage due to sub-threshold porosity $\phi$ also evolves.
In the formulation, the implicit void volume fraction $\phi$ is related to the void concentration $\eta$ and the average void size $\nu$ by $\phi \equiv \eta \nu / ( 1 + \eta \nu )$.
To model nucleation, the void concentration $\eta$, a number density, evolves according to:
\begin{equation} \label{eq:etadot}
\dot{\eta} = \left( N_1 \left(\frac{2^2}{3^3} - \frac{J_3^2}{J_2^3}\right) + N_3 \frac{p}{\sigma_\text{vm}} \right) \eta \, \dot{\epsilon}_p 
\end{equation}
where $p = 1/3 \, \sigmab\cdot\Ib$ is the pressure, $J_2 = 1/2 \tr \sb^2 = 1/3 \, \sigma_\text{vm}^2$ and $J_3 = 1/3 \tr \sb^3$.
The $N_1$ component responds to shear, and the $N_3$ component corresponds to nucleation due to triaxiality $p/\sigma_\text{vm}$.
The evolution relation for $\phi$, 
\begin{equation} \label{eq:phidot}
\dot{\phi} = \sqrt{\frac{2}{3}}\dot{\epsilon}_\text{p}\frac{1 - (1-\phi)^{m+1}}{(1-\phi)^{m}}\sinh\left(\frac{2(2m-1)}{2m+1}\frac{p}{\sigma_\text{vm}} \right) + (1-\phi)^2 \dot{\eta}\nu_0 \ ,
\end{equation}
includes nucleation and growth of voids, where the damage exponent $m$, and the volume of newly nucleated voids $\nu_0$, are additional model parameters.
Lastly, once the void fraction $\phi$ exceeds a threshold $\phi_\text{max}$ the material, as discretized by finite elements, is considered completely failed and acts as a pore.

The calibrated parameters are summarized in \tref{tab:parameters} and further details can be found in \cref{khalil2019modeling}.

\begin{table}[t]
\centering
\begin{tabular}{|lc|c|}
\hline
Parameter & & Value \\
\hline
Young's modulus (GPa)         & $E$       & 240 \\
Poisson's ratio               & $v$       & 0.27 \\
Yield strength (MPa)          & $Y$       & 600   \\
Initial void size ($\mu$m$^3$)    & $\nu_0$   & 0.1 \\
Initial void density ($\mu$m$^{-3}$) & $\eta_0$  & 0.001  \\
Flow exponent                 & $n$        & 10 \\
Damage exponent               & $m$        & 2 \\
Flow coefficient              & $f$        & 10        \\
Isotropic dynamic recovery    & $R$        & 4\\
Isotropic hardening (GPa)     & $H$        & 5\\
Initial hardening (MPa)       & $\kappa_0$ & 460\\
Shear nucleation              & $N_1$      & 10      \\
Triaxiality nucleation        & $N_3$      & 13\\
Maximum damage                & $\phi_\text{max}$  & 0.5  \\
\hline
\end{tabular}
\caption{Constitutive model parameters.}
\label{tab:parameters}
\end{table}

\subsection{Model geometry} \label{sec:geometry}
\begin{figure}
\centering
\includegraphics[width=0.16\textwidth]{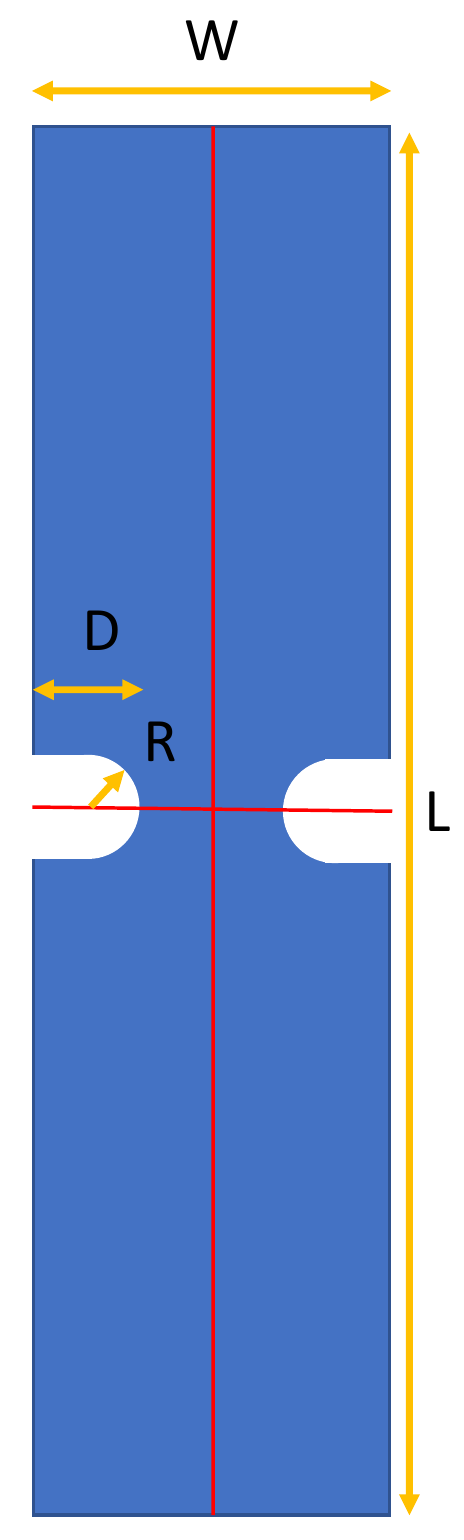}
\caption{Schematic of notch geometry.}
\label{fig:schematic}
\end{figure}

As discussed in the Introduction, we selected the double U-notched tension specimen shown in \fref{fig:schematic} since this a common geometry for mechanical tests \cite{boyce2013morphology,furnish2016fatigue,cook2019predicting} and it allows us to investigate the effects of stress concentration with a few geometric parameters, namely the notch depth $D$ and notch radius $R$.
The overall dimensions of all the specimens were fixed at 1$\times$1$\times$4 mm$^3$, as was element size ($\approx$ 0.05 mm).
As discussed in \cref{khalil2019modeling}, fixing the element size across all the mesh realizations ameliorates the impact of mesh dependence due to softening on the quantities of interest.
We varied the notch radius $R$ in the range 0.05--0.45 mm and depth $D$ over the range 0.15--0.45 mm to create a set of nominal geometries.
As will be seen in the results in \sref{sec:results}, notch radius largely determines the height of the potential localization zone and the acuteness of the stress concentration, whereas notch depth influences the width of the potential localization zone.

\begin{figure}
\centering
\subfloat[$R=$ 0.15 mm, $D=$ 0.05 mm]
{\includegraphics[width=0.2\textwidth]{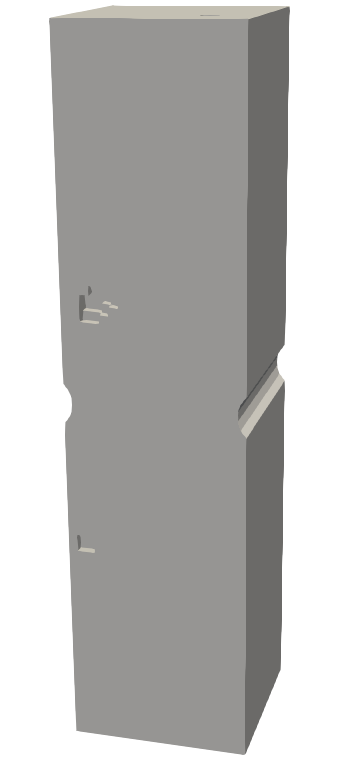}}
\subfloat[$R=$ 0.45 mm, $D=$ 0.05 mm]
{\includegraphics[width=0.2\textwidth]{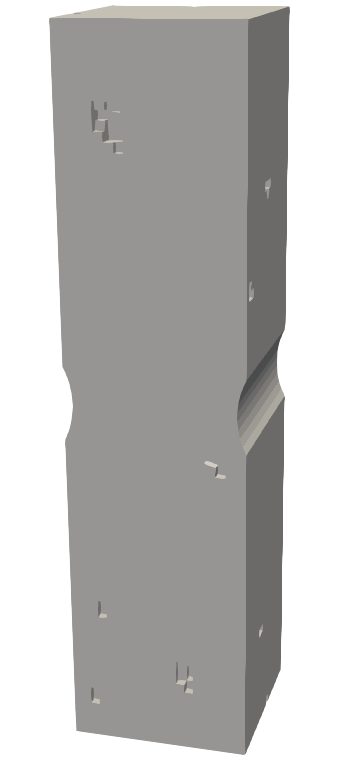}}
\subfloat[$R=$ 5 mm, $D=$ 0.05 mm]
{\includegraphics[width=0.2\textwidth]{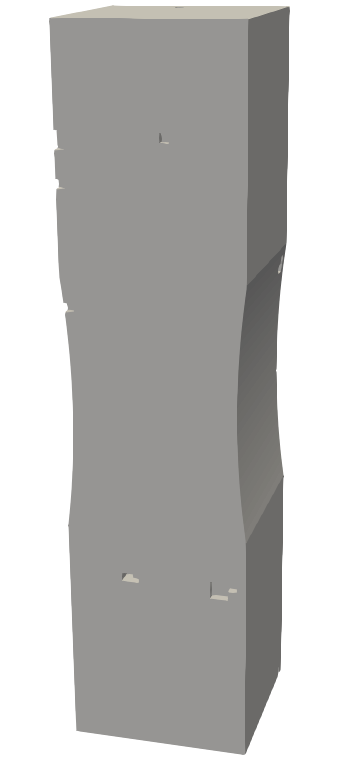}}
\caption{Representative realizations of notched geometries with explicit pores
}
\label{fig:geometry}
\end{figure}

Explicit pores were added to each of the nominal geometries.
We generated realizations of the explicit porosity using a Karhunen–Lo\`{e}ve process that was tuned to reproduce the experimentally observed average porosity $\bar{\phi}$ = 0.008 and spatial correlation length $\approx$ 0.05~mm as described in detail in \cref{khalil2019modeling}.
The largest pore in the ensemble was 81 elements (mean radius of 0.13~mm), and the average pore size was 4.1 elements (mean radius of 0.05 mm) across 10,000 realizations. 
Since this process was applied to the mesh, the smallest explicit pores were one element.
Pores were allowed reside at the surface of the nominal geometry and acted as surface imperfections.
\fref{fig:schematic} shows the nominal geometry and \fref{fig:geometry} shows sample realizations of the explicit porosity on variations in nominal notch geometry. 

Tensile loading was effected with minimal displacement boundary conditions on the top and bottom surfaces (refer to \fref{fig:schematic}) such that these boundary surfaces remain planar but can latterly expand or contract freely.
The displacement loading rate was held fixed at 8~$\mu$m/s (0.002 1/s strain rate).

A number of dimensionless scales are apparent in this model. 
The analytic expressions for stress concentration provide a few based on the nominal geometry, shown in  \fref{fig:schematic}, such as the gauge ratio $D/W$ and notch aspect ratio $R/D$.
The size of the explicit pores relative to the geometry of the notch suggest others such as the ratio of the notch radius $R$ to that of the largest pore.

\begin{figure}
\centering
\includegraphics[width=0.6\textwidth]{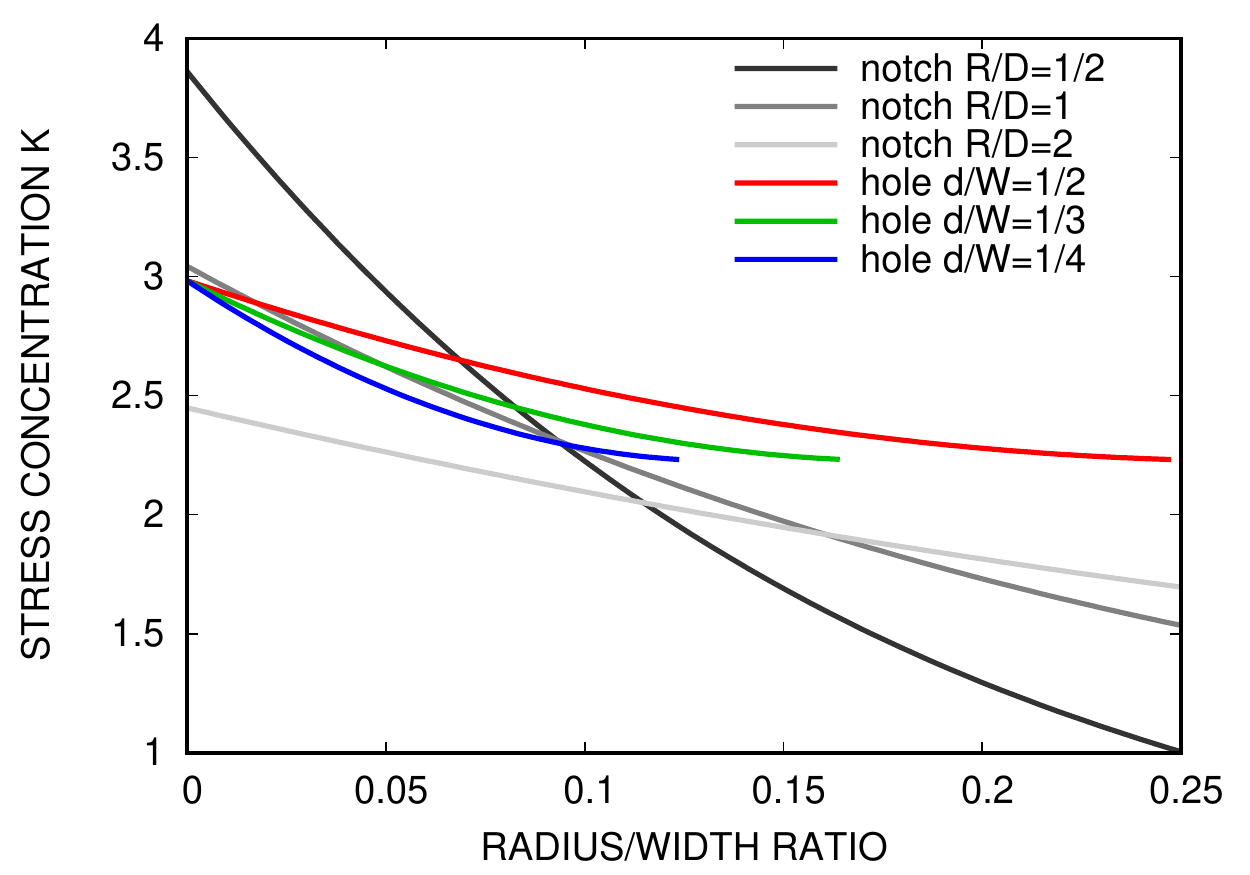}
\caption{Elastic stress concentrations due to holes (at various locations $d/W$) and notches (of various aspect ratios $R/D$) in a thin plate.
Horizontal axis is radius of the U-shaped notch $R$ or cylindrical pore  $r$ relative to the width of the specimen $W$.
}
\label{fig:hole-notch}
\end{figure}

The stress concentration $K \equiv {\sigma_\text{max}}/{\bar{\sigma}}$ induced by notches of this type have been estimated analytically \cite[Chart 2.4]{pilkey1997petersons} and takes the form:
\begin{equation}
K_\text{notch} = K_\text{notch}(R/W,D/R) 
\end{equation}
where $\sigma_\text{max}$ is the maximum stress, $\bar{\sigma}$ is the nominal stress (without the notch), and $W$ is the width of the specimen.
Our voids are coarsely discretized and three dimensional, nevertheless the stress concentration due to cylindrical holes \cite[Chart 4.3]{pilkey1997petersons} is relevant as an indication of scaling of the stress concentration.
These stress concentrations are of the form
\begin{equation}
K_\text{void} = K_\text{void}(r/W,d/r))
\end{equation}
where $r$ is radius of the hole, $d$ is the distance from the hole to the surface, and $W$ is the width of the rectangular specimen.
Following Fig. 3.10 in  \cref{schijve2001fatigue} for semicircular double notches, \fref{fig:hole-notch} shows the competing effects of U-notches and sub-surface pores.
Taken qualitatively, this analysis shows that sharp, deep notches $R/D \ll 1$ will be the dominant effect in a porous specimen with small pores; however, there is a cross-over point where a large pore far from the surface can dominate.
Conversely, pores play a dominant role as stress concentrations in the regime where the notch is wide and shallow $R/D \gg 1$.
Although not quantitatively predictive for the simulations where strong pore-pore effects are at play, these observations do provide expectations that bear out in our simulations discussed in the next section.
While not present in the analytical solutions, the length of the specimen $L$ relative to the dimensions of the notch also plays a role, albeit a less significant one, as the simulations discussed in \sref{sec:simulation} will illustrate.

\clearpage

\section{Plasticity-Damage-Porosity Simulations}\label{sec:results}

With the computational model described in \sref{sec:simulation}, we ran 1000--10,000 realizations of nominal geometries with different notch depth $D$ and notch radius $R$ to explore the effects of specific porosity configurations on failure.
The simulations display a complex interplay between explicit pores and stress concentrations caused by the geometry.
We isolate physical effects by:
changing notch dimensions to alter the stress concentrations and necking zone; and including or omitting the explicit porosity which engenders random small scale stress concentrations.
The behavior of the ensemble of realizations is described in this section and a descriptive statistical model of the behavior in the subsequent section.

\subsection{Rectangular specimen}
\begin{figure}
\centering
{\includegraphics[width=0.5\textwidth]{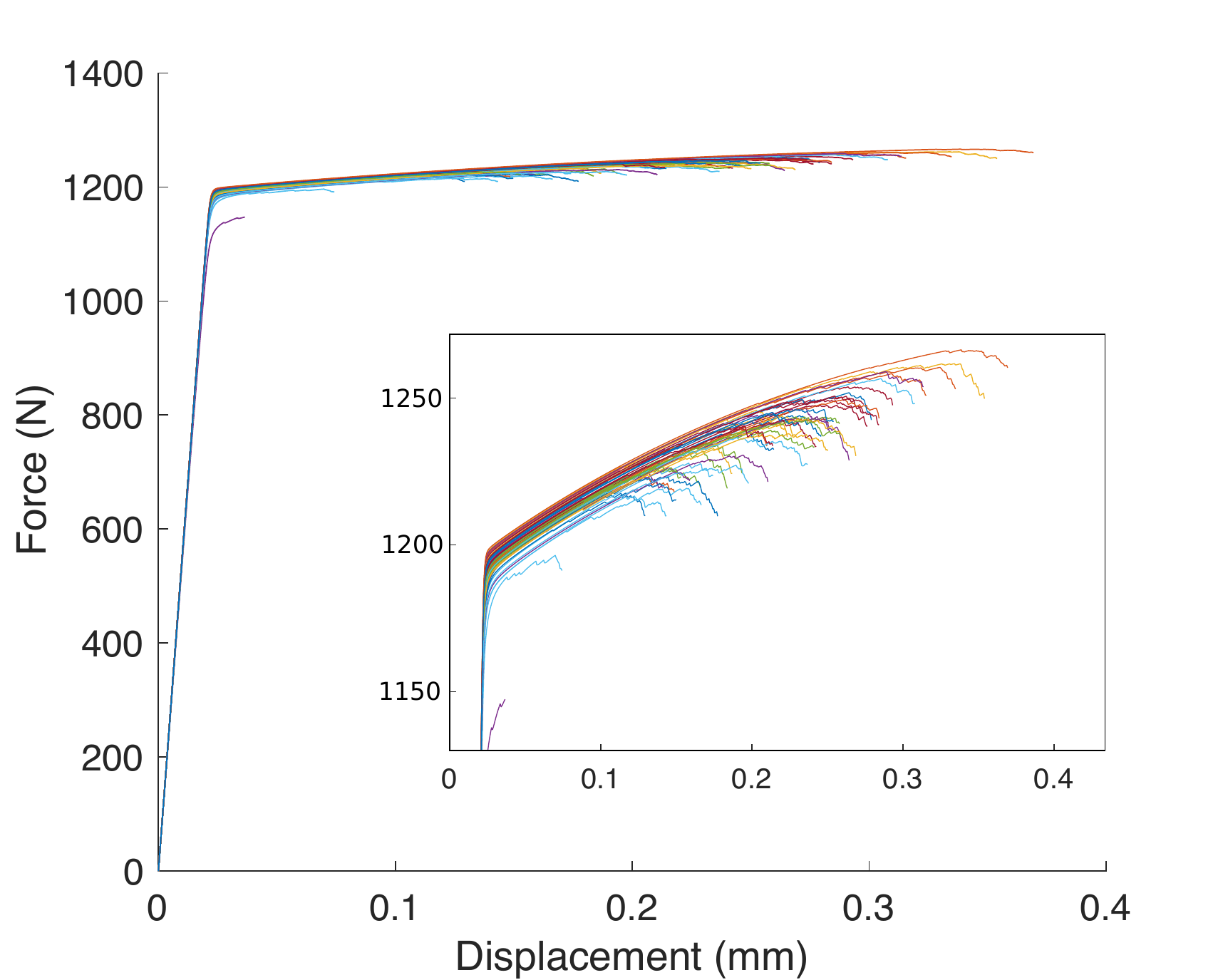}}
\caption{Force-displacement curves for 50 realizations of porosity in un-notched, rectangular specimens. The inset shows a zoomed-in portion of the force-displacement plot to better indicate where failure occurs.
}
\label{fig:nonotch_force_displacement}
\end{figure}

To determine the baseline effect of pores without stress concentrations we ran 10,000 explicit porosity realizations of the nominal geometry without the double U-notch.
\fref{fig:nonotch_force_displacement} illustrates the variation of specimen stress-strain response due to variations in 50 of the explicit porosity realizations.
Given that the average porosity is fixed, the elastic response shows little variability across the realizations.
Nevertheless there are outliers. 
In fact, one rectangular specimen with several large surface voids failed at a noticeably lower force and displacement than the others shown.
There is some variation in yield and hardening among the other specimens, but the dominant variability is in failure strain.
In our previous work \cite{khalil2019modeling} we explored yield strength, ultimate strength and several quantities related to failure as variable features of the stress-stain response.
Here we focus on the failure location as the main quantity of interest in order to understand the phenomenology driving the variability in failure strain.
The failure location can be identified by a localized region of highly damaged material in the model.
This localization is precipitated by a rapid, approximately exponential rise in the damage field culminating in a cascade of complete element failures as element damage reaches the specified threshold of $\phi_\mathrm{max}$ nearly simultaneously across the region.
At the final loading step, we identify this spike in damage by locating all elements with damage values within 10\% of the specimen's maximum damage.
We then define the failure location as the average of the $z$-coordinate values of these highly damaged elements.
This threshold based metric provided a more complete and sharper picture of the failure surface than considering only elements that met the maximum damage $\phi_\text{max}$ criterion.

\begin{figure}
\centering
\includegraphics[width=0.5\textwidth]{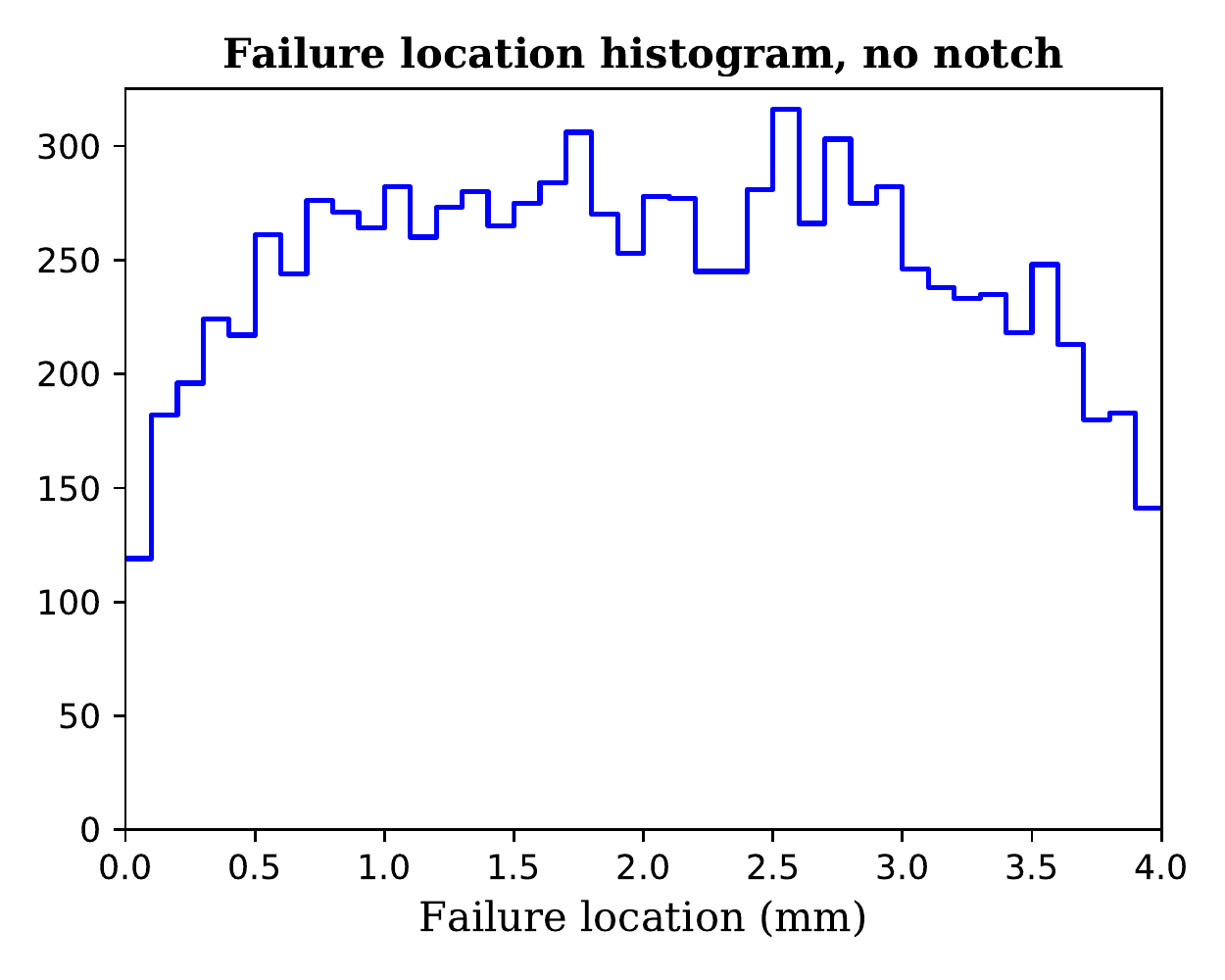}
\caption{Histogram of failure locations with no notch for approximately 10,000 realizations of explicit porosity.
}
\label{fig:fail_locations_noNotch}
\end{figure}

\begin{figure}
\centering
\includegraphics[width=0.9\textwidth]{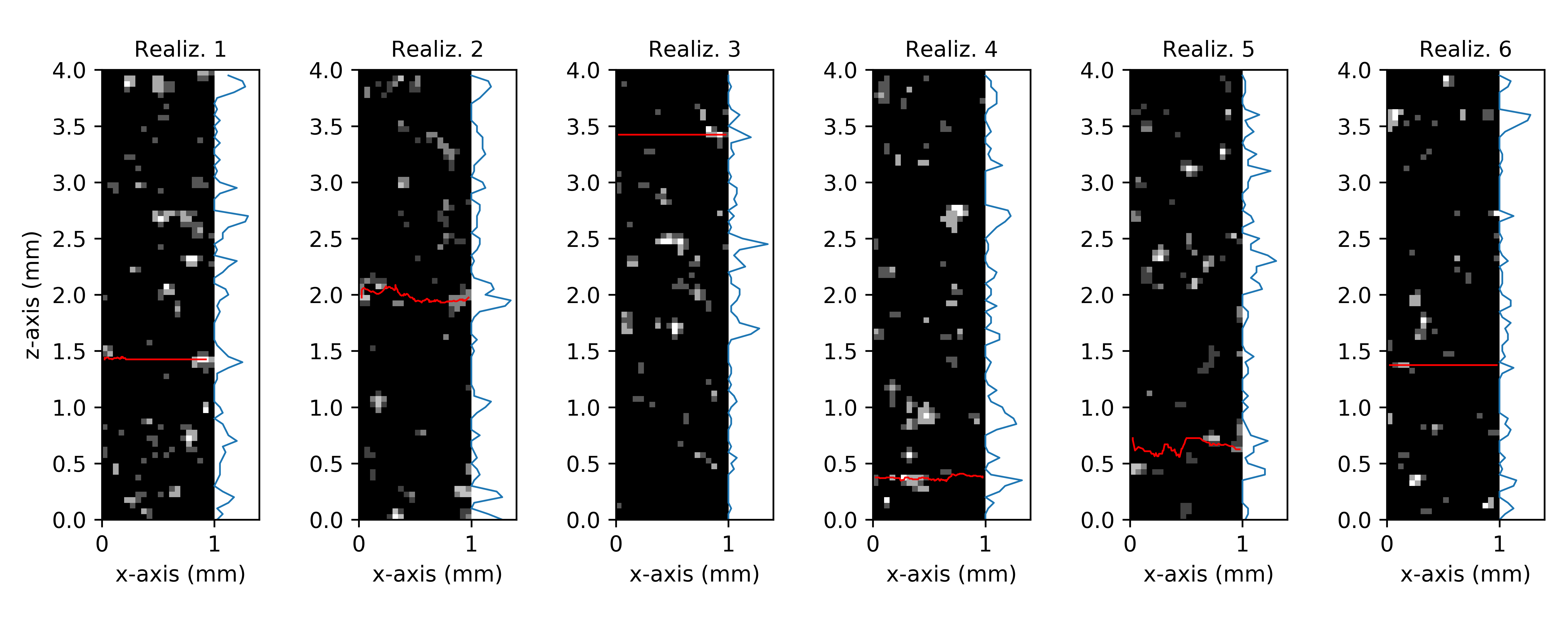}
\caption{Failure surface and explicit pores for six different realizations of explicit pores with no notch. The failure surface is represented by the red line and is constructed from a rolling average along the $x$-axis of the location of high-damage elements. The high-damage elements are correlated with the failure location. The grayscale heatmap represents the 2D distribution of explicit pores, with the number of void voxels summed along the $y$-axis. The blue histograms shows the distribution of voids along the $z$-axis.
}
\label{fig:correlation_noNotch}
\end{figure}

\begin{figure}
\centering
\subfloat[Realization 2]
{\includegraphics[width=0.7\textwidth]{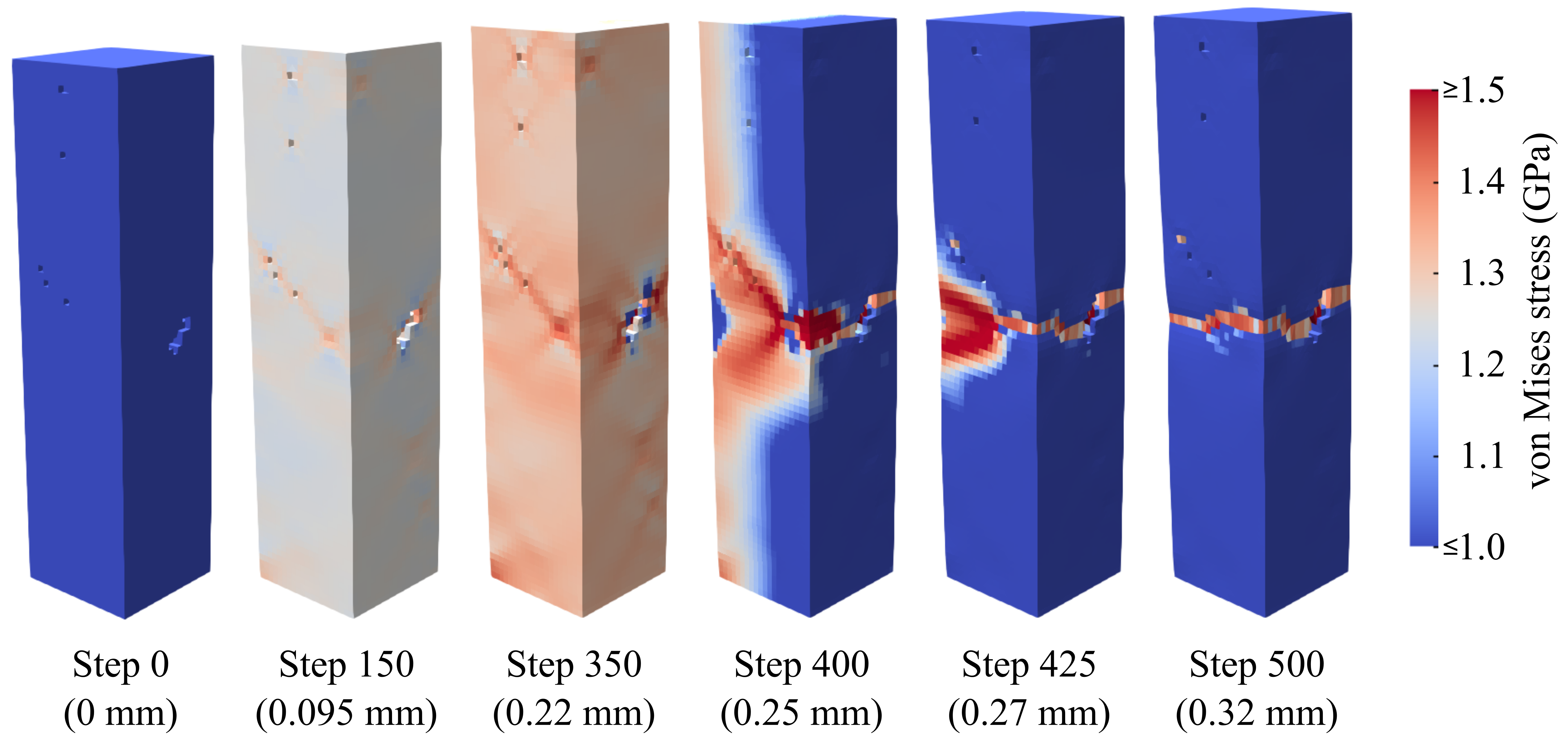}}

\subfloat[Realization 6]
{\includegraphics[width=0.7\textwidth]{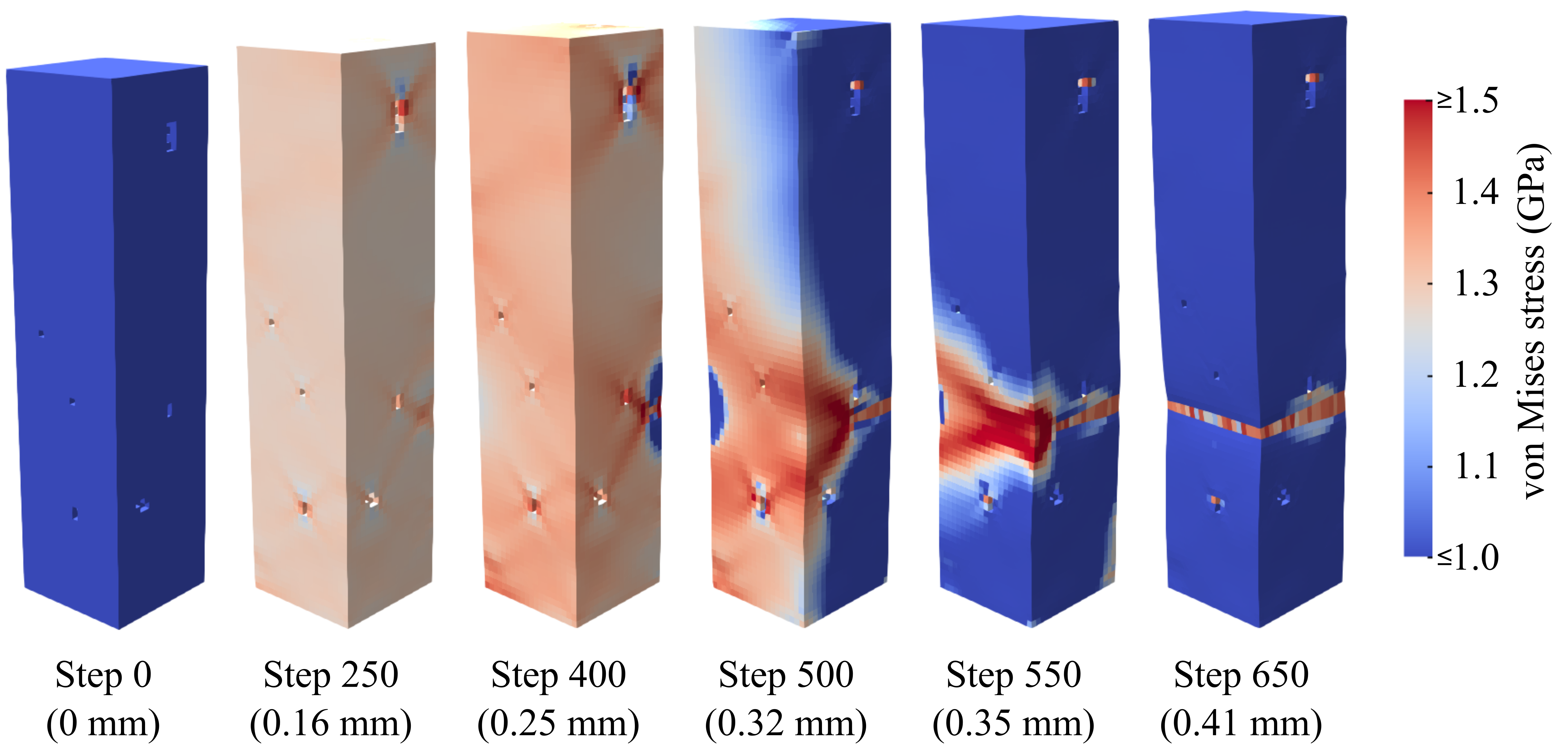}}
\caption{Evolution of the von Mises stress field for two realizations where (a) failure occurs at the largest void concentration, and (b) failure does not occur at the largest void concentration (compare with Fig. \ref{fig:correlation_noNotch}).
The total displacement of the top surface at each step is shown in parentheses.
}
\label{fig:damage_noNotch}
\end{figure}

\fref{fig:fail_locations_noNotch} indicates that without a notch there is a nearly uniform distribution in failure location across the specimens with slight but apparent boundary effects.
These are effects are evidently due to a reduction of spatial-correlation effects for pores near the boundaries, which are subject to kinematic constraints that enforce that the specimen ends remain planar and parallel.
We found that the failure locations were not highly correlated with any obvious aspect of the porosity field, \eg the location of the largest void or the void closest to the surface.
In fact, the correlation coefficient of the location of the largest void and failure was only 0.26.
In \fref{fig:correlation_noNotch}, the failure locations have been superimposed on the explicit porosity as seen through the thickness of the specimen, clearly illustrating the difficulty with simple correlative analyses.
While individual specimens are seen to fail near large pores, in regions with multiple pores, or where pores are adjacent to the specimen surface, detailed statistical analysis reveals that none of these observations have significant explanatory power in a statistical sense.
Furthermore, analysis of the local fields in the specimens does not offer a clear picture of any qualitative differences among the failure phenomenology in different specimens.
For example, \fref{fig:damage_noNotch} shows the failure modes for a realization that fails at the largest concentrations of pores and one that does not.
It is evident that the interactions among the pores are strong and multiplex, and that the stress fields are qualitatively similar from one realization to the next.
However, there is little to suggest a simple mechanistic explanation for why one realization fails where pore concentration is high, and the other does not.
The evident difficulty in devising a simple mechanistic predictor of failure location motivates the statistical treatment developed in \sref{sec:model}.

\clearpage

\subsection{Notched specimens} \label{sec:notched}

In this section, we survey the influence of varying the nominal geometric features of the U-notch.
In contrast to the stochastic nature of the failure processes observed in the nominally uniform specimens of the previous section, we expect to observe effectively deterministic effects of stress concentration due to the double U notch on the failure location.
We explored a wide range of the geometric notch parameters $R$ and $D$, from a true {\it notch} $R/D \ll 1$ with a significant stress concentration factor to a {\it taper} $R/D \gg 1$  that only slightly focuses the stress field.
As mentioned in \sref{sec:geometry}, the radius of the smallest notches we examine is comparable in size to the largest explicit voids.

\begin{figure}
\centering
{\includegraphics[width=0.9\textwidth]{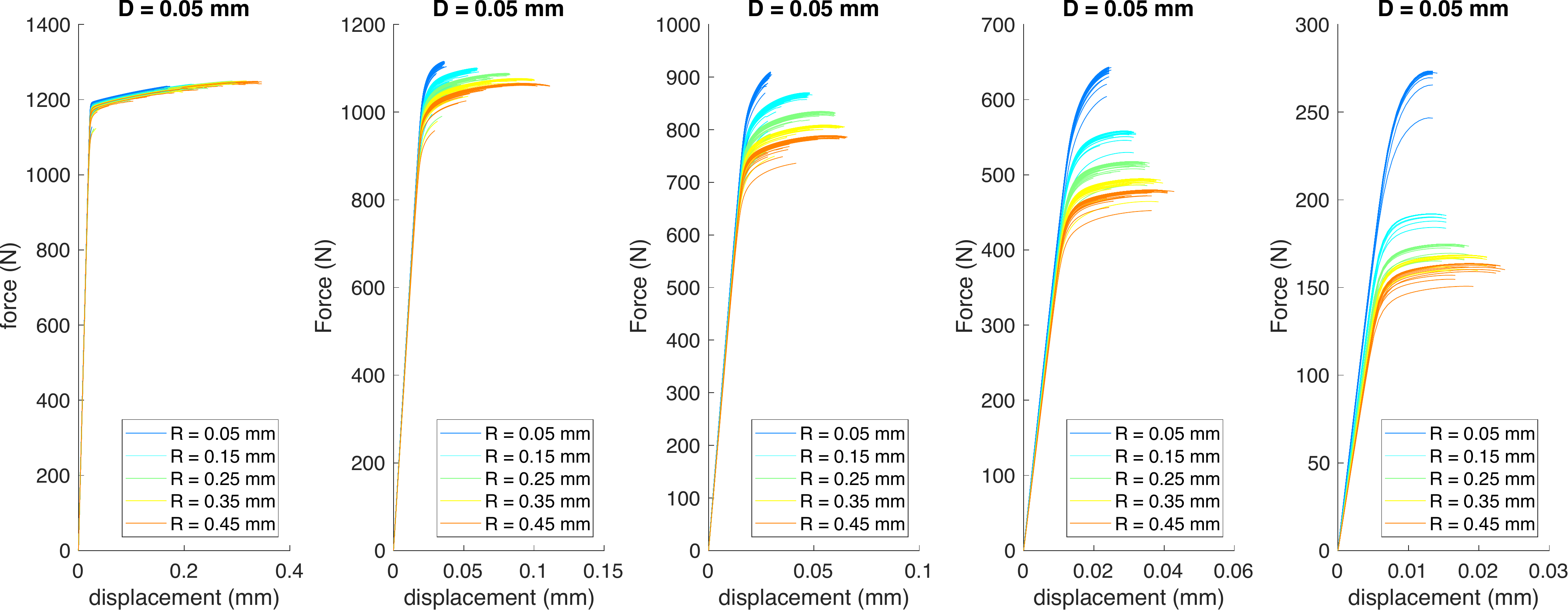}}
\caption{Force-displacement curves for range of  $R$ and $D$.
Note the range of vertical and horizontal axis changes across the panels.
}
\label{fig:force_displacement_data}
\end{figure}

\fref{fig:force_displacement_data} shows the force-displacement response of the ensemble of specimens.
The response of shallowest notches resembles those of the specimens without notches (\fref{fig:nonotch_force_displacement}) in that there is minimal variation except in the failure strain.
A trend to fail earlier with decreasing notch radius (increasing stress concentration and triaxiality) is apparent, and this trend is evident for all notch depths.
With increasing notch depth, in addition to experiencing more rapid damage evolution, specimens with sharper notches also experience an apparent delay in the onset of yield due to the higher triaxiality of the stress field and some apparent hardening due to a transition toward lower triaxiality during concentrated plastic deformation.
For deeper notches, yield and hardening respond more strongly to the particular arrangement of pores in the realization as the cross-sectional area of the effective gauge section decreases and approaches the average pore size.
In general, it is apparent that the notch depth sets the minimum cross-section and hence the mean maximum stress and width of the plastic zone.
The notch radius controls the intensity and spatial extent of the stress concentration and hence the height and concentration of the plastic zone and the number of voids experiencing these conditions.

\begin{figure}
\centering
\includegraphics[width=\textwidth]{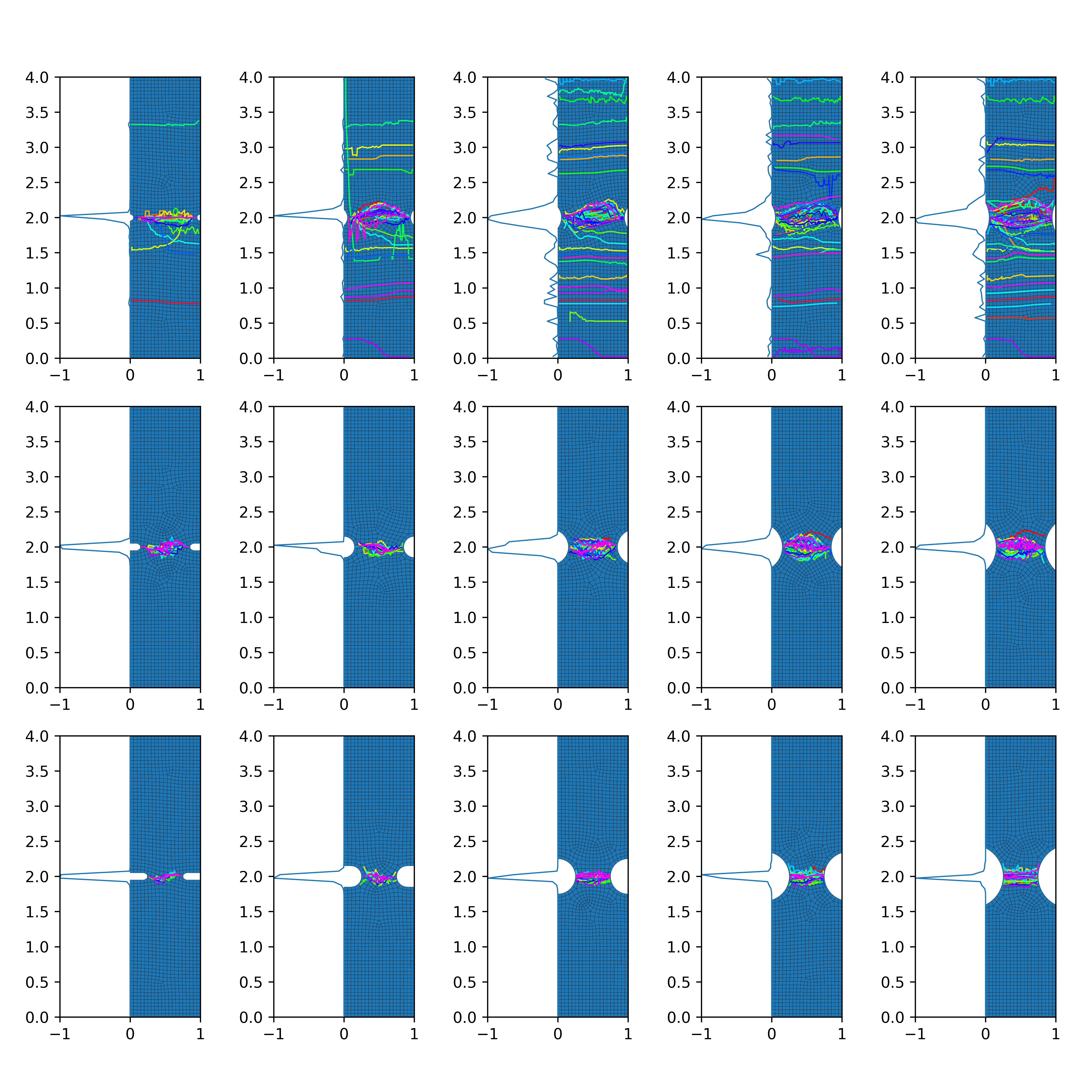}
\caption{Plot presenting the range of failure surfaces for 50 explicit void realizations for each notch geometry. Each colored line represents a different realization and is constructed from a rolling average along the x-axis of the location of high-damage elements. The high-damage elements are correlated with the failure location. The histogram in each plot shows the distribution of high-damage elements across all realizations for each geometry.
}
\label{fig:fail_geom}
\end{figure}

To understand the competition of the nominal geometry and material porosity in driving failure, we use distributions of failure location as surrogates.
Intuitively, high variance in the failure location indicates a dominance of the random porosity, while low variance indicates a dominance of the nominal geometry.
\fref{fig:fail_geom} shows the distribution of the failure locations across realizations for a sampling of the nominal geometries.
For most geometries the distribution of failure is centered at the mid-plane with a roughly lenticular locus.
The failures that lie outside this region tend to be horizontally oriented and not spatially clustered.
Consistent with the interpretation of the analytic stress concentrations illustrated in \fref{fig:hole-notch},
we found that only the shallowest notches had a significant likelihood of failure outside the vicinity of the notch.
In fact, \fref{fig:fail_geom} demonstrates that deep and/or narrow notches with high stress concentration have little uncertainty in failure location.
This analysis suggests that the critical parameter for determining the influence of the material porosity on failure strain is the ratio of the notch depth to the average pore size, which is an indicator of the relative stress concentrations as well as the extent of the region of increased stress.
If this parameter is much less than unity, the material porosity is significant influence on failure location and should be considered wherever failure analysis is required.
For larger values of the ratio of the notch depth to the average pore size, we will show analysis based solely on the stress concentration of the notches is sufficient.

\begin{figure}
\centering
\includegraphics[width=\textwidth]{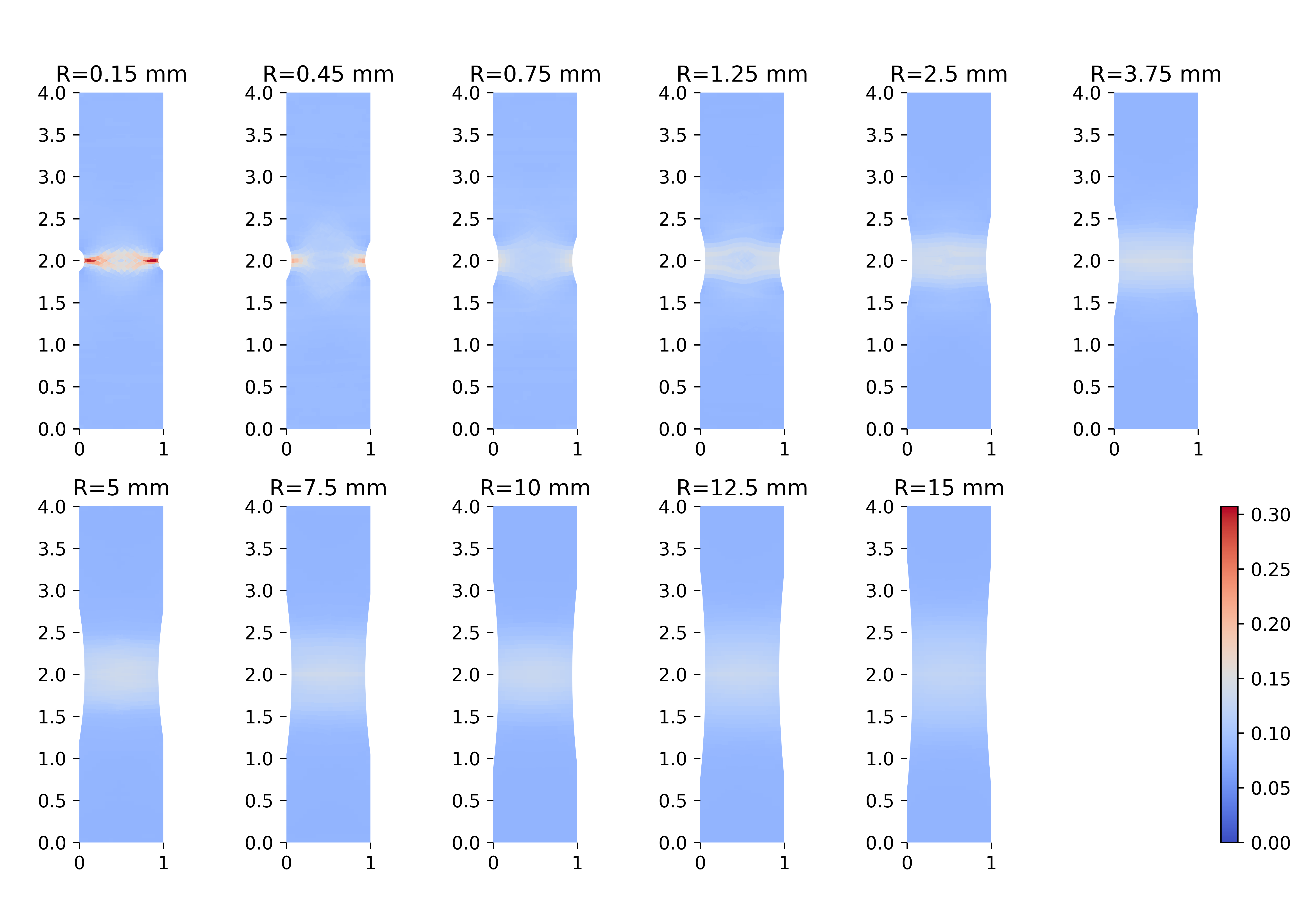}
\caption{Ensemble averages of the final damage field for notches with various radii ($D=$ 0.05 mm), averaged over approximately 1,000 realizations of explicit porosity. The damage field was mirror-symmetrized about the horizontal midline of the specimen to mitigate slight biases in the results due to asymmetries in the top and bottom halves of the meshes.
}
\label{fig:avg_dmg}
\end{figure}

\clearpage 

\begin{figure}
\centering
\includegraphics[width=\textwidth]{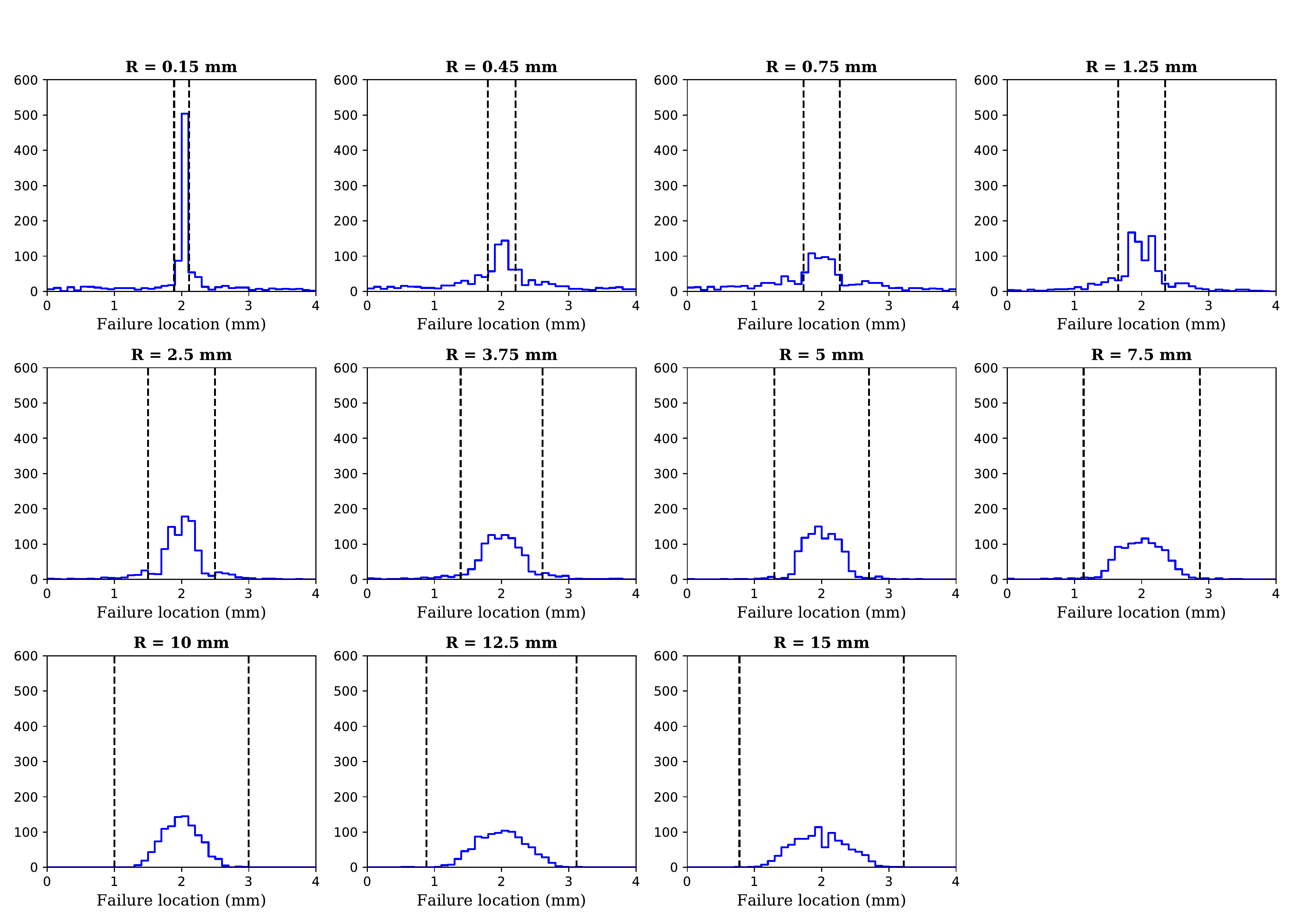}
\caption{Histograms of failure locations with a notch depth of 0.05 mm and various notch radii. Each geometry has approximately 1,000 realizations of explicit porosity. The dashed lines mark the edge of the notch.
}
\label{fig:hist_notch}
\end{figure}

We now shift focus to specimens with shallow notches/tapers where the failure location is more unpredictable.
For the collection with $D=$ 0.05~mm, \fref{fig:avg_dmg} shows the ensemble averaged damage fields after failure has occurred for various geometries and how the localized zone responds to the width of the notch.
For sharp (small $R$) notches the damage zone is localized. 
As the notches widen so does the damage zone but only up to a point, at which the zone remains effectively constant width. 
This can also be observed in the progression of histograms shown in \fref{fig:hist_notch}.
The failures are distributed over the specimen for the notches with smaller radii, but are generally contained within the damage zone as the radii get larger, $R \ge 7.5$ mm.

\fref{fig:hist_notch} shows the reduction of failure locations to a one-dimensional distribution along the long/$z$ coordinate of the specimens.
Generally speaking the distribution is narrow and peaked for the narrow notches and correspondingly broad for the wide notches.
However, for the narrowest notches there is a significant background distribution of failures outside the notch.
As with the damage field for the broadest notches the distribution width does not directly scale with the width of the notch and eventually remains essentially constant for the widest notches/tapers.
In fact the transformation of the distributions displays a complex dependence on the geometric factor $R$. 
For intermediate notch widths the central peak superimposed on a fairly uniform background transitions to a tri-modal distribution for intermediate notch radii.
For the widest notches the distribution is bell-shaped with no failures outside the notch. 

We can interpret this behavior in the context of energy dissipation.
If the notch creates only a narrow zone of damage, the power due to external loading needs to dissipate outside this zone, and does so by creating localized failure outside the notch. 
However, a wider notch creates a larger damage zone that is sufficient to  dissipate the external power supplied by the boundary loading. 
Consequently, on average there is no excess of power to be dissipated in failures outside this zone, and therefore they are unlikely to arise. 
In this case, the localization stabilizes in a constant width zone sufficient to dissipate the prescribed external power.

\begin{figure}
\centering
\subfloat[pressure]
{\includegraphics[width=0.4\textwidth]{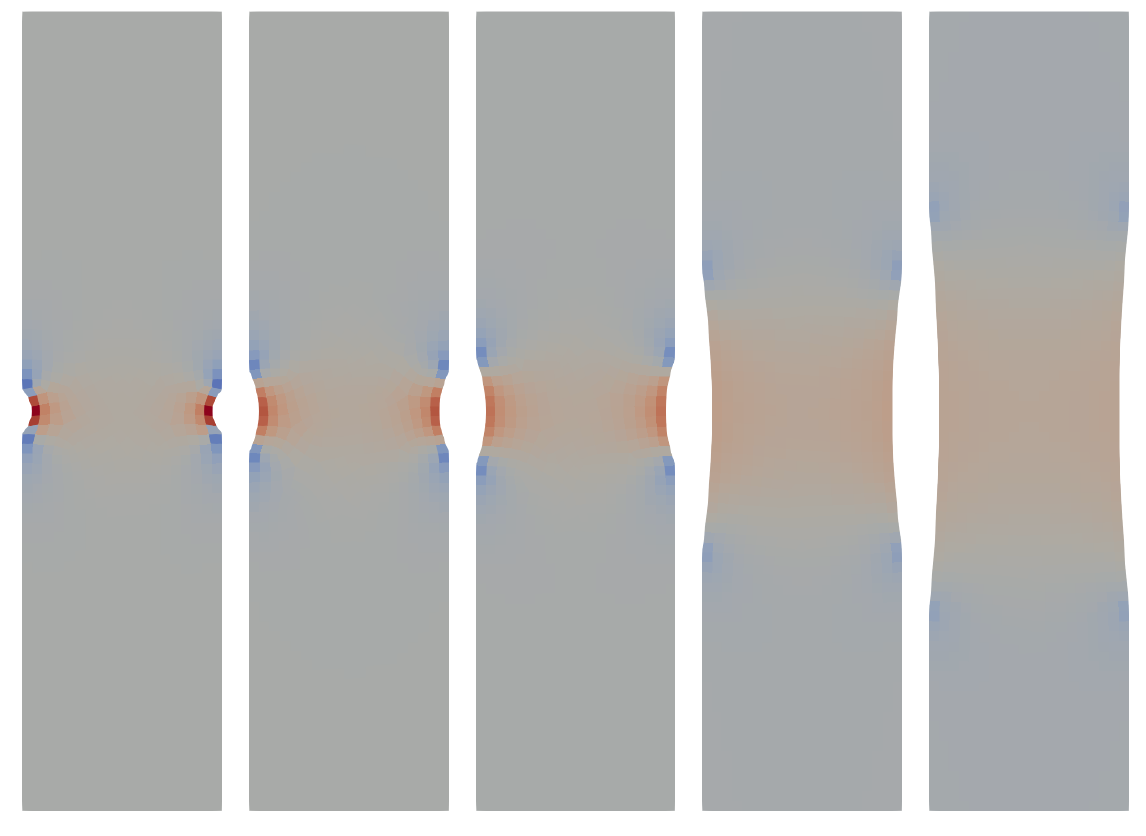}\label{fig:driving_fields:pressure}}

\subfloat[von Mises stress]
{\includegraphics[width=0.4\textwidth]{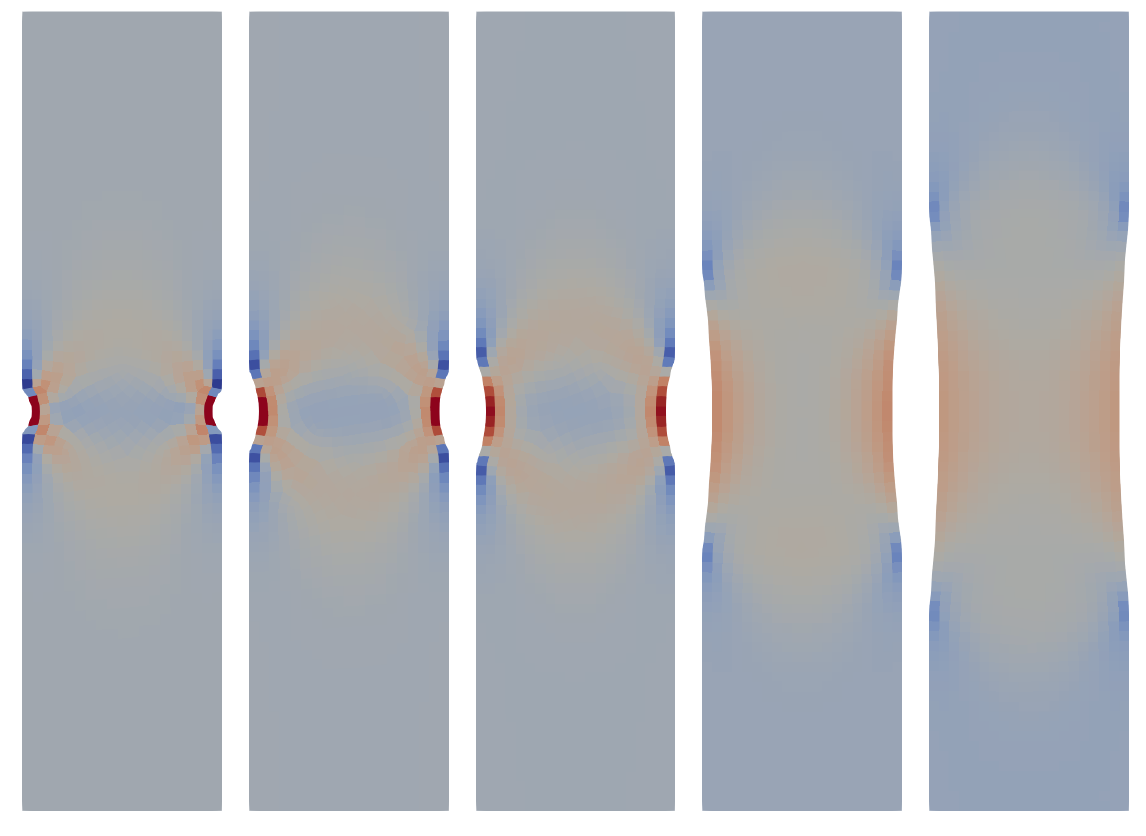}\label{fig:driving_fields:vonmises}}

\subfloat[damage]
{\includegraphics[width=0.4\textwidth]{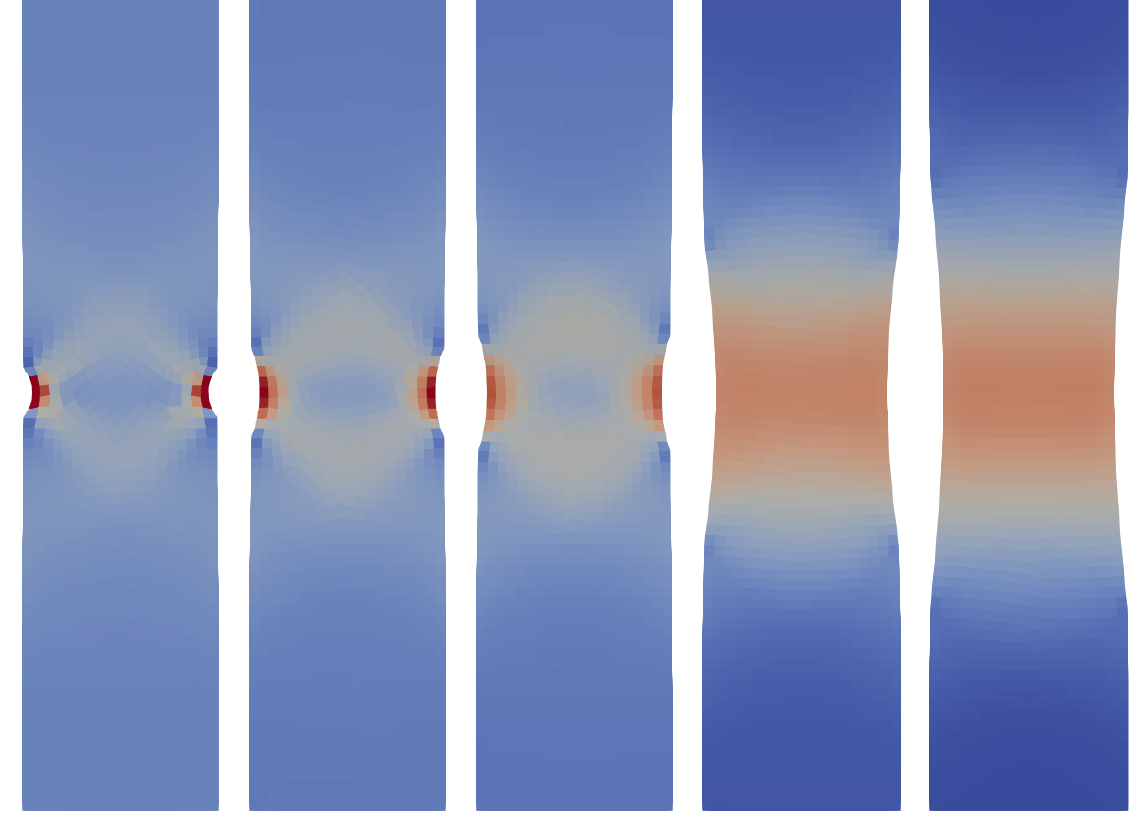}\label{fig:driving_fields:damage}}

\caption{Fields without explicit pores for $R$ = 0.15, 0.45, 0.75, 5.0, 10.0 mm at 4.7\% strain. 
For pressure: 1.1 MPa (blue), 4.9 MPa (red);
for von Mises stress: 6.6 MPa (blue), 11.5 MPa (red);
and for damage: 0.08 (blue), 0.10 (red).
}
\label{fig:driving_fields}
\end{figure}

\begin{figure}
\centering
\subfloat[mean, minimum and maximum stress]
{\includegraphics[width=0.45\textwidth]{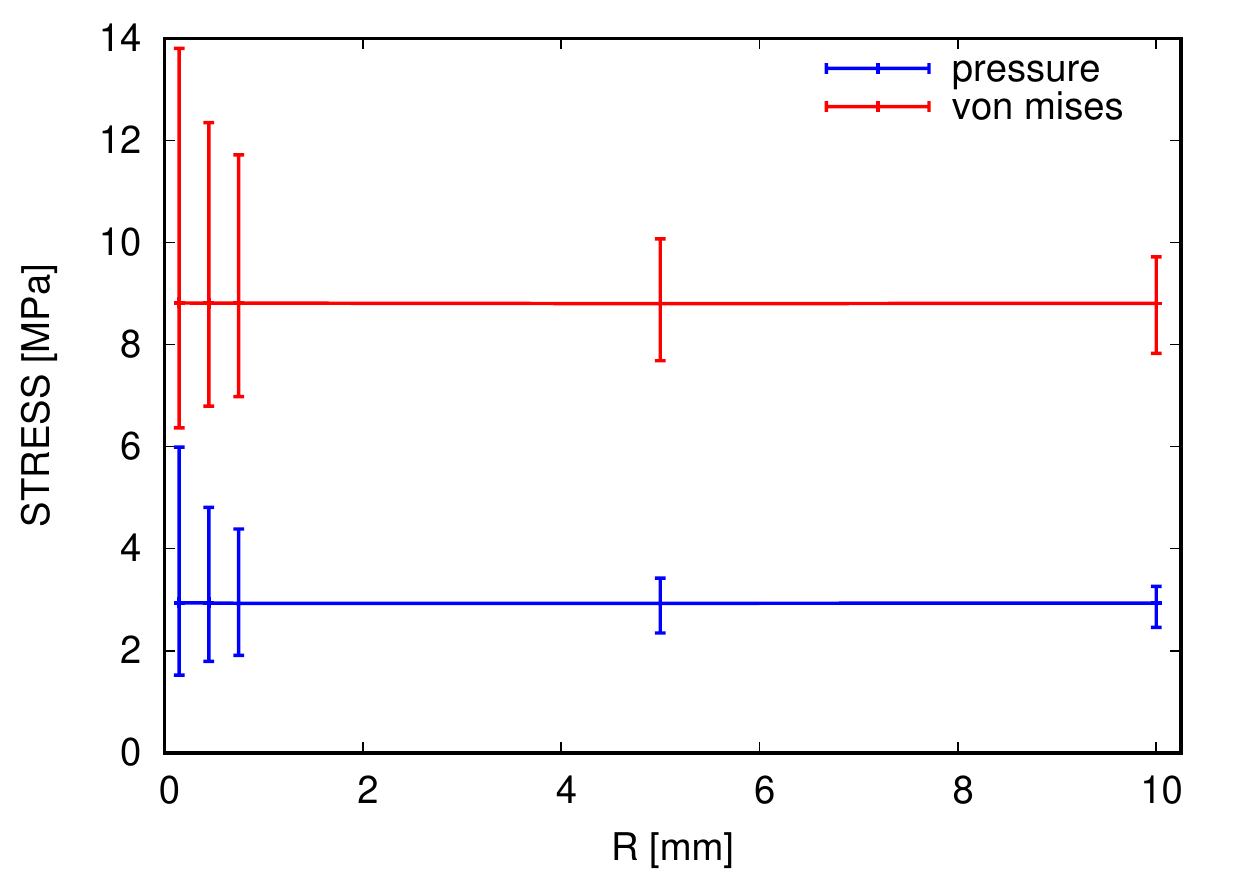}}
\subfloat[stress concentration]
{\includegraphics[width=0.45\textwidth]{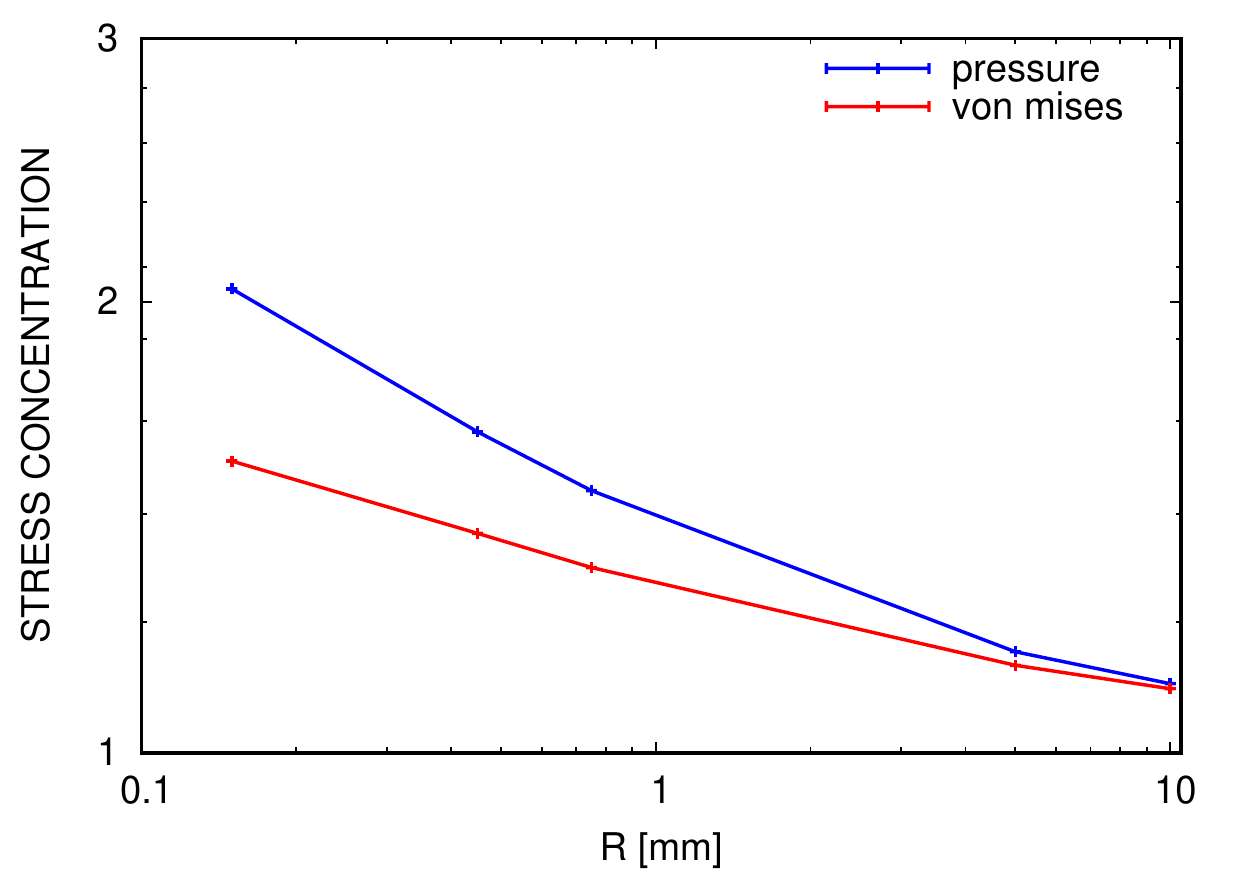}}
\caption{Stress variance and concentration for the pressure $p$ and von Mises stress $\sigma_\text{vm}$.
The stress concentration for the von Mises stress is approximately $\log K =  0.0069 \log^2 R  -0.0870 \log R +  0.2602$ and for the pressure $\log K =  0.0187 \log^2 R  -0.1530 \log R +  0.3572$.
}
\label{fig:stress_concentration}
\end{figure}

Clearly, by adding the notch we impose a background stress field that makes failure more likely at certain pores.
Given the non-linearity of the model the effect is not strictly a superposition of the fields due to the notch and to the porosity, as assumed in classical analytical models.
Nevertheless it is illuminating to try to decompose the effects by simulating the same nominal geometry without pores.
\fref{fig:driving_fields} illustrates how the pressure and von Mises stress change with geometry (the fields are shown at nominal strain of 4.7\%).
It is apparent that the damage distribution at this stage of the deformation resembles the von Mises stress field for narrow notches but transitions to resemble the diffuse, fairly uniform distribution of pressure for wider notches.
In particular the diamond pattern of the von Mises stress field for the narrow notches appears to be correlated with the damage shown in \fref{fig:avg_dmg} and the failure pattern in \fref{fig:fail_geom}.
This superficial resemblance, while hinting at the complex interactions of plasticity and void evolution represented in models such as \eref{eq:phidot}, is insufficient to explain the complex, history-dependent behavior of ductile materials under low-triaxiality loading.
In fact, the well-established concept of stress concentration, which reduces spatial distributions of stress to a single scalar for failure analysis, tells a conflicting story.
\fref{fig:stress_concentration}a shows the range of magnitudes of the pressure and von Mises stress over the collection of shallow notch geometries.
As expected, the extremes of the stress fields become more pronounced as $R$ decreases.
We can also obtain empirical estimates of stress concentration, plotted in \fref{fig:stress_concentration}b, where it is apparent that even the shallowest notches exhibit noticeable concentrations.
(Note that these empirical stress concentrations are based on the elastic response, post yield these measures can change significantly.)
From this figure, it is clear that the triaxiality (the ratio $p/\sigma_\text{vm}$) increases for smaller $R$, consistent with the reduced ductilities shown in \fref{fig:force_displacement_data}.

Thus, on one hand it is clear that the pressure is important in determining the global performance of the specimen, while local behavior can be more closely correlated with the von Mises stress.
To reconcile these observations, we develop a statistical model for the interaction of the structural geometry-induced stress fields and the material porosity.
We relate the statistical model to design parameters for a particular geometry, demonstrating a practical application of the model; and, we use empirical stress concentrations, such as those plotted in \fref{fig:stress_concentration}b, to link the statistical model back to classical concepts in damage mechanics to establish the generality of the model.


\section{Statistical models of failure locations} \label{sec:model}

Since the explicit pore locations are unlikely to be available for design, we develop statistical models for a probabilistic characterization of failure location.
We reduce the three-dimensional failure location to a one-dimensional random variable representing the axial location of the centroid of the failure region to enable robust statistical analysis with the amount of data collected.
Gaussian mixture models (GMM) are widely-used finite mixture models used in capturing the underlying distribution of observed data in various fields of applications \cite{everitt2014finite,lindsay1995mixture,marin2005bayesian,mclachlan2004finite,titterington1985statistical}. 
However, Gaussian distributions are symmetric and have infinite support.
Here, the failure location, normalized by the specimen length, falls within the unit interval, and its distribution is not generally symmetric, as observed from analysis based on kernel density estimation.
Thus, we employ a mixture of beta distributions, also known as a beta mixture model (BMM) \cite{bouguila2006practical,ma2011bayesian}. 
The BMM, once calibrated with the data provided by the set realizations described in \sref{sec:simulation}, is subsequently used to distinguish between the different sources of failure.
As conveyed in \sref{sec:physical}, we are able to provide an estimate for the likelihood of the failure having been caused or initiated by the presence of a notch, as opposed to being associated with the explicit porosity.

\subsection{Methodology}
Finite mixture models \cite{mclachlan2004finite} utilize a mixture of parametric distributions to capture the scatter in the observed data. Although we are dealing with uni-variate data, namely the failure location along the length of the specimen for this application, this analysis can be directly extended to model multivariate data, such as the full three dimensional failure location.

\begin{figure}
\centering
\includegraphics[width=0.5\textwidth]{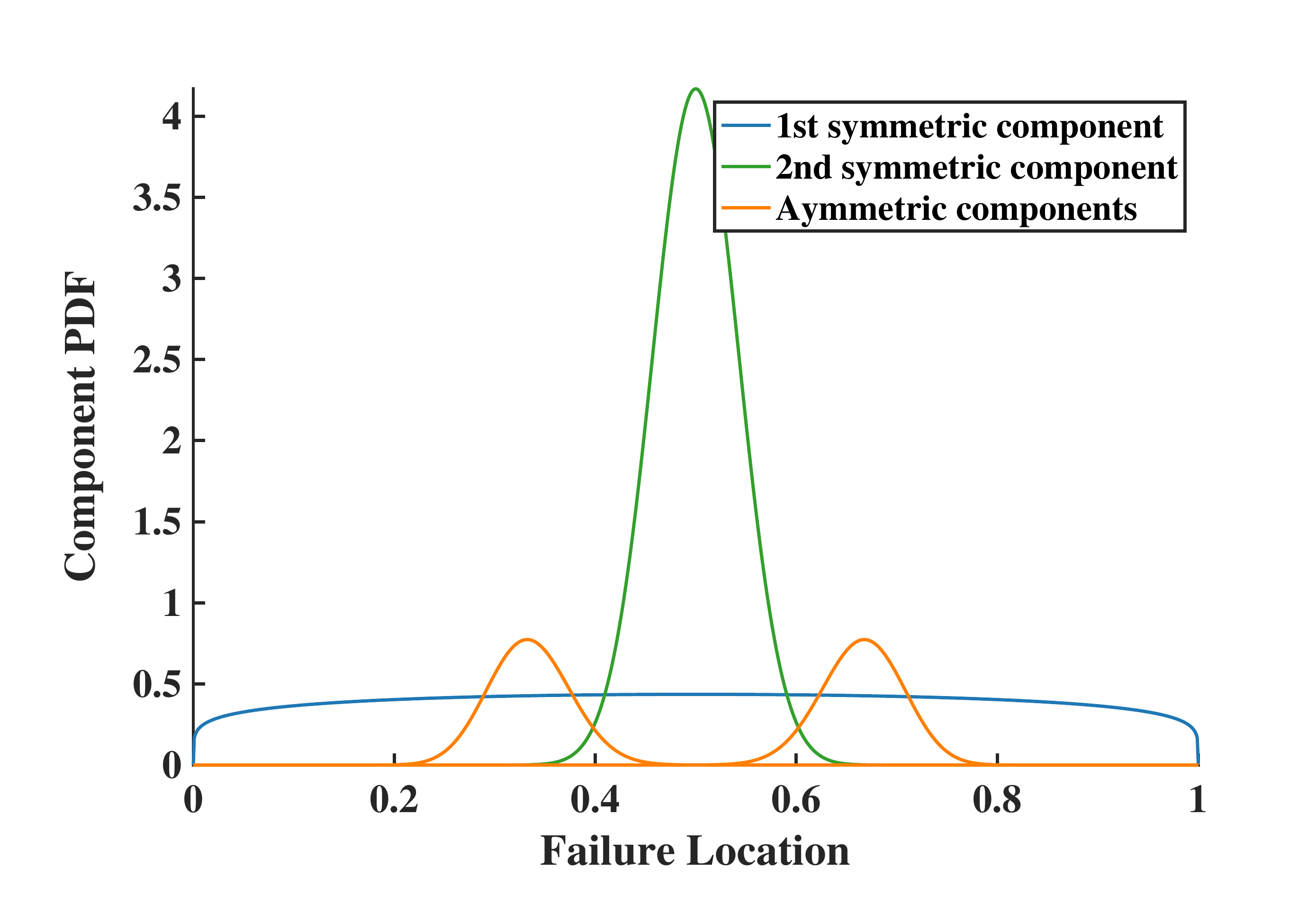}

\caption{Calibrated Beta mixture model components for the notch of radius $R = 0.75$ mm, refer to \fref{fig:fitted_BMM}d. We associate the broad first component with the background density of pores, the narrow second component with stress concentration at the deepest part of the notches, and the combined asymmetric/off-center components with the inclined, shear-dominant concentrations emanating from the notches.}
\label{fig:fitted_BMM_components}
\end{figure}

Denoting by $z$ the failure location normalized by $L=4$ mm to the unit interval, we define the $K$-component uni-variate BMM as
\begin{equation} \label{eq:f}
    f(z \ ; {\bf W },{\bf A},{\bf B}) = \sum_{k=1}^{K}w_k \ {\rm Beta}(z \ ;a_{k}, b_{k}) \ ,
\end{equation}
where
\begin{equation}
    {\bf W } = \{w_1, \ldots, w_K \},\ {\bf A} = \{{\bf a}_1, \ldots, {\bf a}_K\} ,\ {\bf B} = \{{\bf b}_1, \ldots, {\bf b}_K\} 
\end{equation}
denote the collection of component weights (or component coefficients), $w_{k}$, and beta PDF hyperparameters (or component parameters) $a_{k}$ and $b_{k}$. 
A beta PDF is given by
\begin{equation}
    {\rm Beta}(z; a, b) = \frac{1}{ {\rm beta}(a, b)} z^{a-1} (1-z)^{b-1},\ a, b >0 \ ,
\end{equation}
where ${\rm beta}(a, b)=\Gamma(a)\Gamma(b)/\Gamma(a+b)$ is the beta function, and $\Gamma(\cdot)$ is the gamma function. 
The shape of the beta distribution depends on two hyperparameters, $a$ and $b$, which determine whether the distribution is symmetric or highly skewed.
See \fref{fig:fitted_BMM_components} for an example.

The calibration of the BMM of the $z$ failure locations reduces to estimating {\bf W}, {\bf A} and {\bf B}, subject to the normalization constraint $\sum_{i=1}^K w_k = 1$.
We handle this constraint by  explicitly expressing the one coefficient  $w_K = 1 - \sum_{i=1}^{K-1} w_k$ in terms of the others for $ K > 1$  and $w_1 = 1$ for $K = 1$, therefore reducing the number of unknowns by 1 coefficient. 
We will estimate the unknown parameters from the available $n \approx 10^5$ independent samples of failure location, $z_j$, $j = 1, \hdots, n$, using the maximum likelihood estimation (MLE) technique \cite{eliason1993maximum,fan1998local}, given by
\begin{align} \label{MLE}
    \left\{ {\bf W, A, B } \right\}_{\rm MLE} & = \argmax_{{\bf W, A, B }} \mathcal{L} \left( {\bf W}, {\bf A}, {\bf B} \ ; z_1 \hdots, z_n \right) \nonumber \\
    & = \argmax_{{\bf W, A, B }} \prod_{j = 1}^n \mathcal{L} \left( {\bf W}, {\bf A}, {\bf B} \ ; z_j  \right) \nonumber \\
    & = \argmax_{{\bf W, A, B }} \prod_{j = 1}^n f(z_j \ ; {\bf W },{\bf A},{\bf B}) \ ,
\end{align}
or, equivalently, through maximizing the log-likelihood given by
\begin{align}
    \left\{ {\bf W, A, B } \right\}_{\rm MLE} & = \argmax_{{\bf W, A, B }} \sum_{j = 1}^n {\rm log} \ f(z_j \ ; {\bf W },{\bf A},{\bf B}) \ ,
\end{align}
where $f$ is given by \eref{eq:f}.

In this context of calibrating a beta mixture model, MLE generally results in a non-concave optimization problem with several local maxima \cite{mclachlan2004finite}. We tackle this optimization issue with a multi-start approach \cite{schoen1991stochastic} whereby the Nelder-Mead simplex method \cite{lagarias1998convergence} is run several times (100 in this investigation) with random starting values for the parameters. 
Another challenge with the BMM arises when dealing with the  $z$ samples that are 0 or 1 in value with corresponding ill-defined likelihood function for certain parameter realizations. 
This is not relevant to this investigation since the $z$ failure locations are strictly greater than zero and less than one with all probability due to the boundary conditions on the tension specimens.

Another practical issue with the use of finite mixture models relates to the choice of number of components, $K$. 
Akaike information criterion (AIC) \cite{akaike1998information} and Bayesian information criterion (BIC) \cite{ghosh2007introduction} have been used to decide the number of components for finite mixture models \cite{roeder1997practical}. 
More generally, Bayesian model selection may also be used, albeit with greater computational demands \cite{corduneanu2001variational}. 
In this investigation, the initial choice of $K$ was informed by AIC and/or BIC for the model selection problem relating to the optimal number of beta components. 
These criteria are readily available since they utilize the likelihood value corresponding to the MLE estimates. 
The number of optimal components varied from 1 to 5 across the various geometries ($R$ in 0.15--15.0 mm) investigated. 
Ultimately, we deviated slightly from the AIC/BIC results in order to arrive at an interpretable representation of the data \cite{gilpin2018explaining} at the expense of a relatively small decrease in data-fit. 
Specifically, we decided to apply AIC to the set of mixture models that are constructed with two symmetric/centered (i.e. $a = b$) components and two asymmetric/off-center (i.e. $a \neq b$) components. 
There is an additional constraint on the two asymmetric/off-center components in that they are mirrored versions of each other, with the weights being equal and the hyper-parameters $a$ and $b$ of the first asymmetric component are equal to $b$ and $a$ of the second component, respectively.
The symmetry of the combined off-center modes is motivated by the data and the boundary value problem.
Therefore, the maximum number of unknowns to be determined through optimization in \eref{MLE} is 6 (4 component parameters and 2 component coefficients/weights), low enough to justify the use of a multi-start globalization scheme of the Nelder-Mead simplex method to solve the associated non-concave problem.

\begin{figure}
\centering
\subfloat[No notch]
{\includegraphics[width=0.30\textwidth]{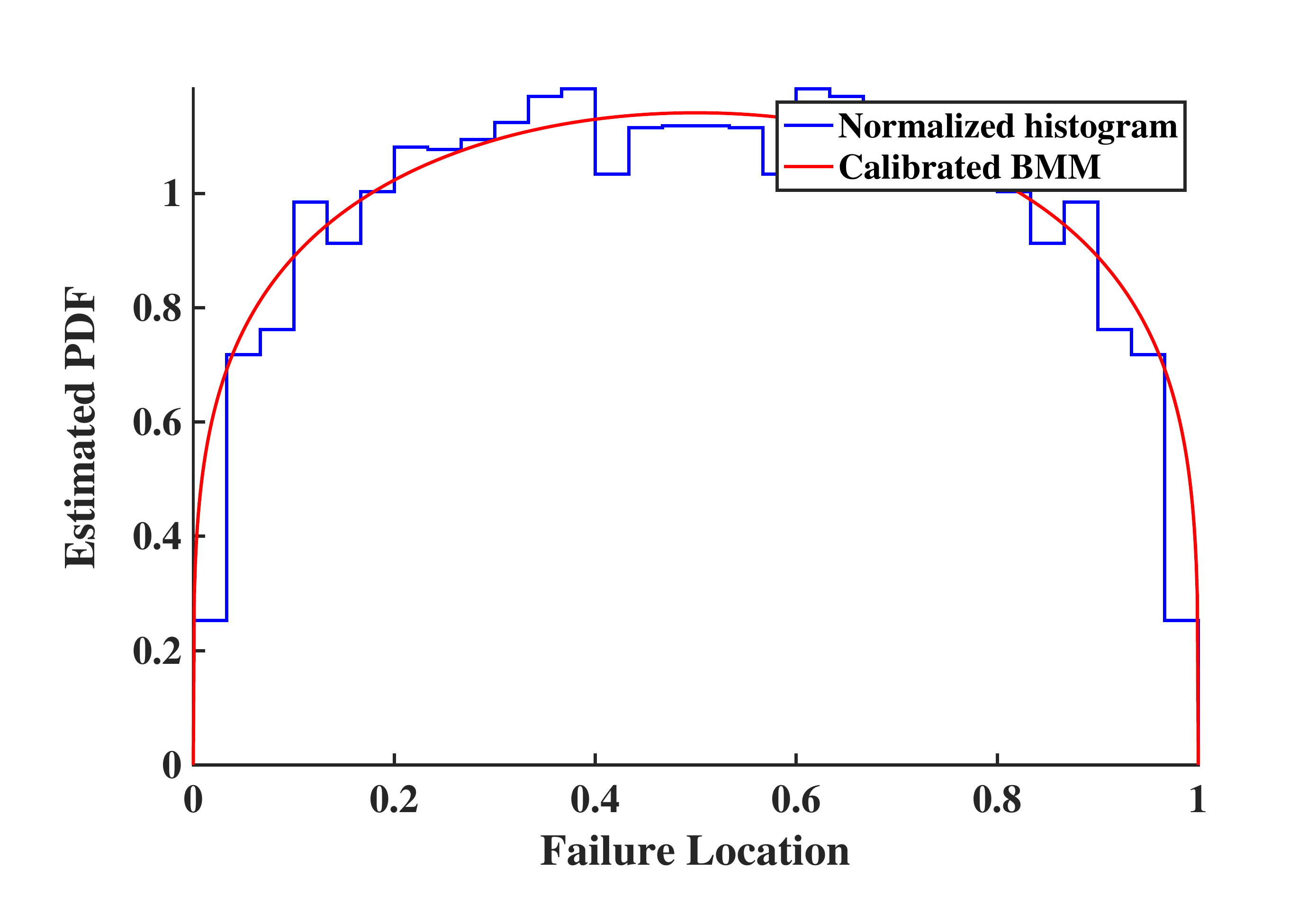}}
\qquad
\subfloat[Notch with 0.15 mm radius]
{\includegraphics[width=0.30\textwidth]{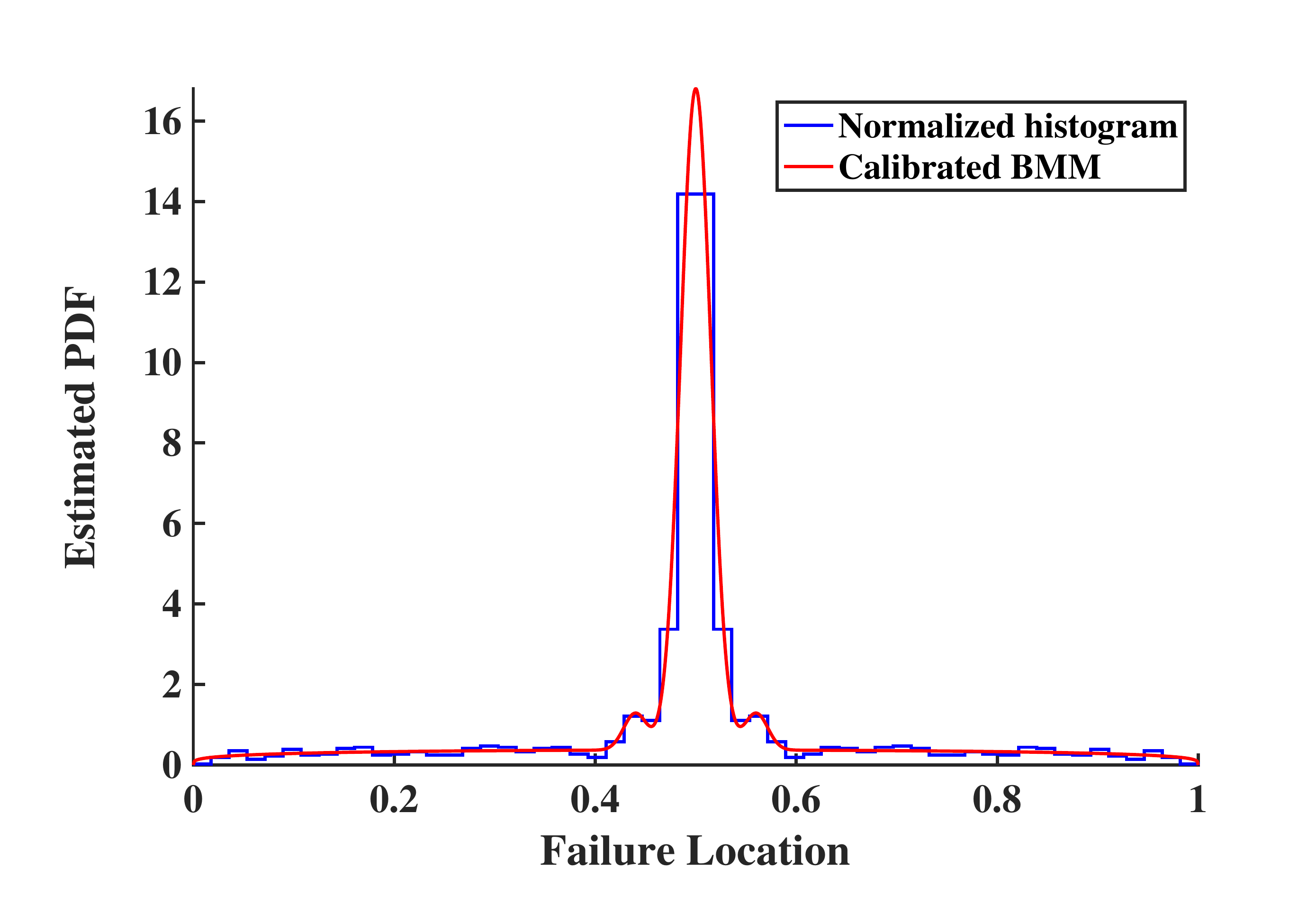}}
\qquad
\subfloat[Notch with 0.45 mm radius]
{\includegraphics[width=0.30\textwidth]{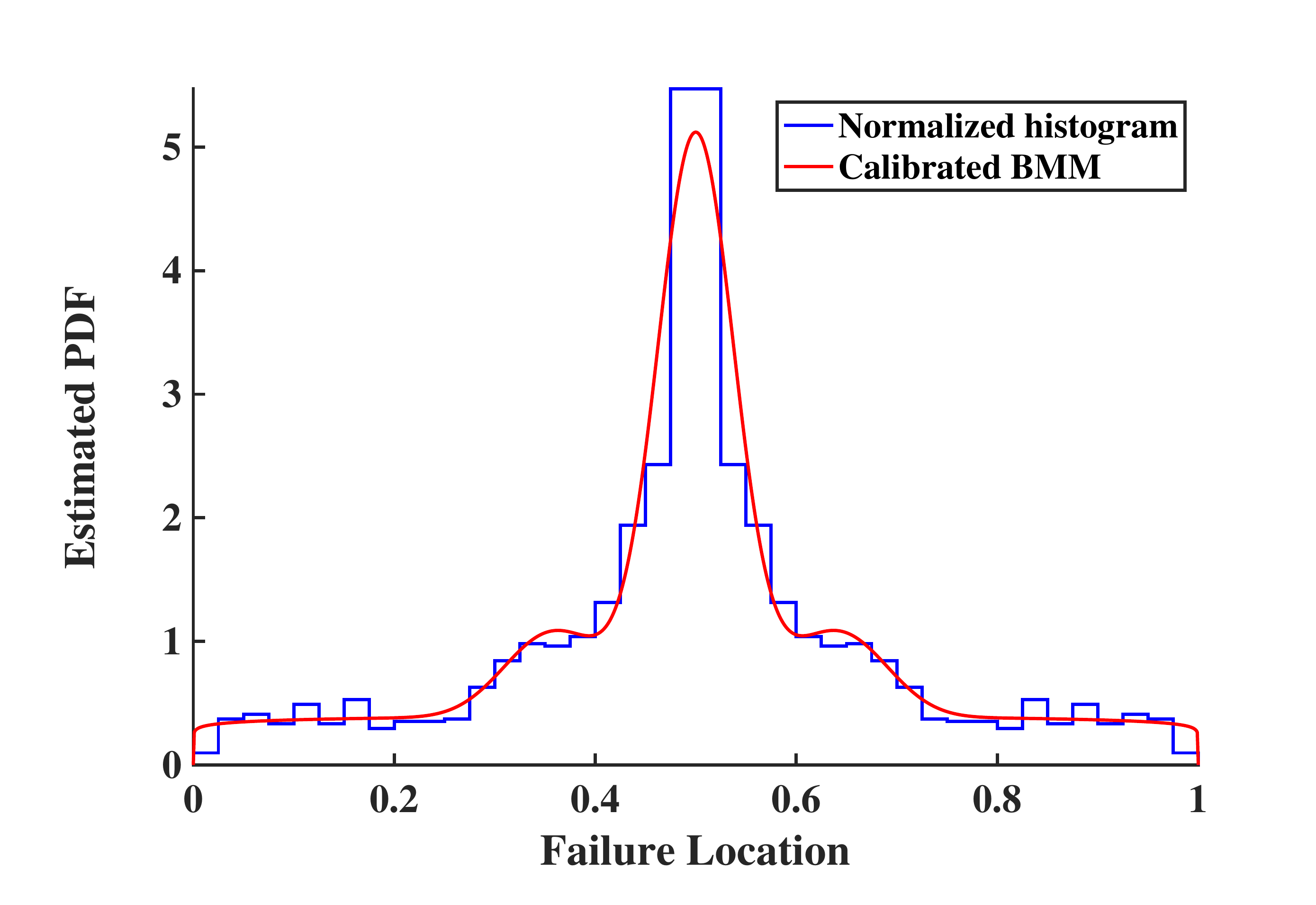}}

\subfloat[Notch with 0.75 mm radius]
{\includegraphics[width=0.30\textwidth]{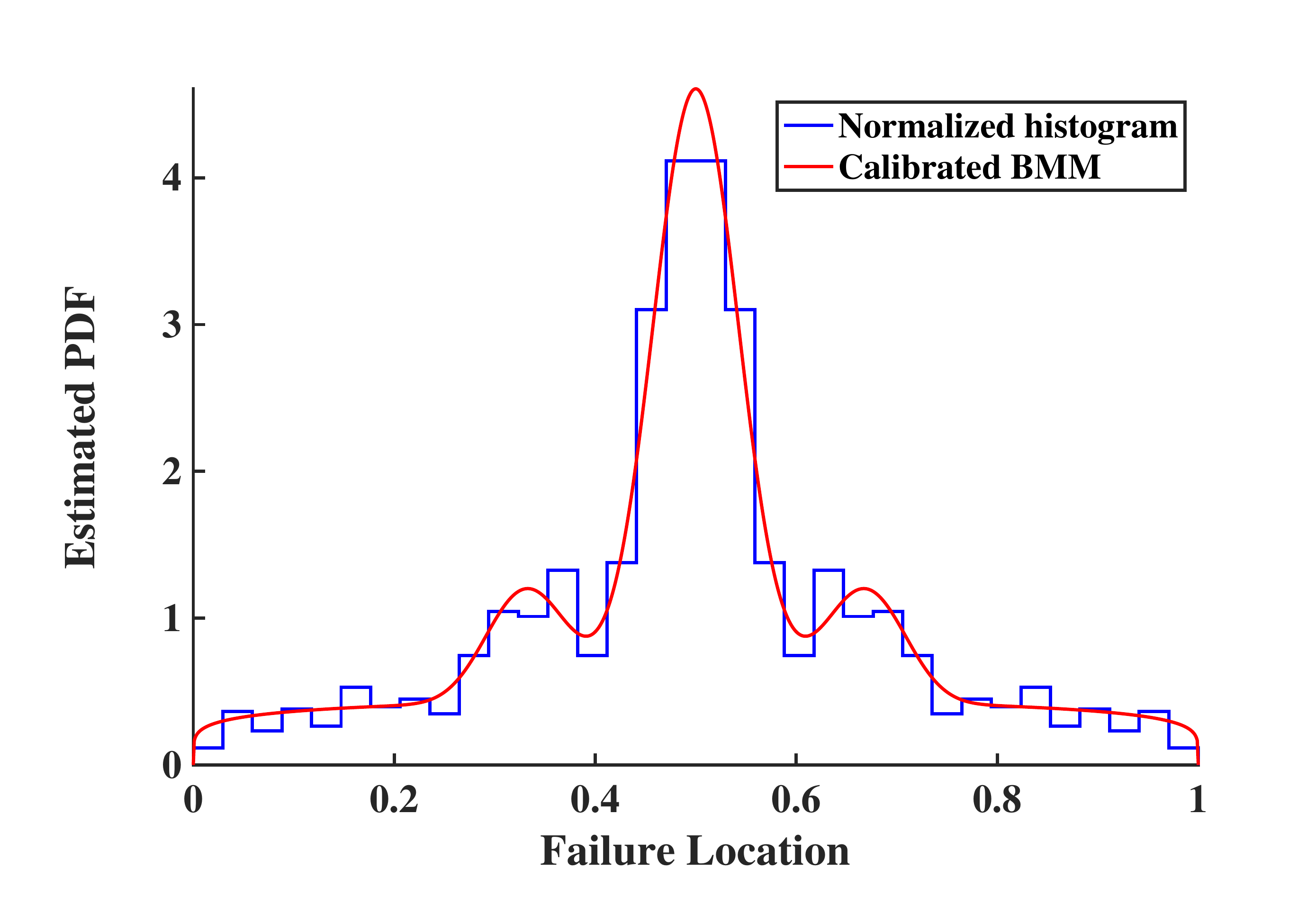}}
\qquad
\subfloat[Notch with 1.25 mm radius]
{\includegraphics[width=0.30\textwidth]{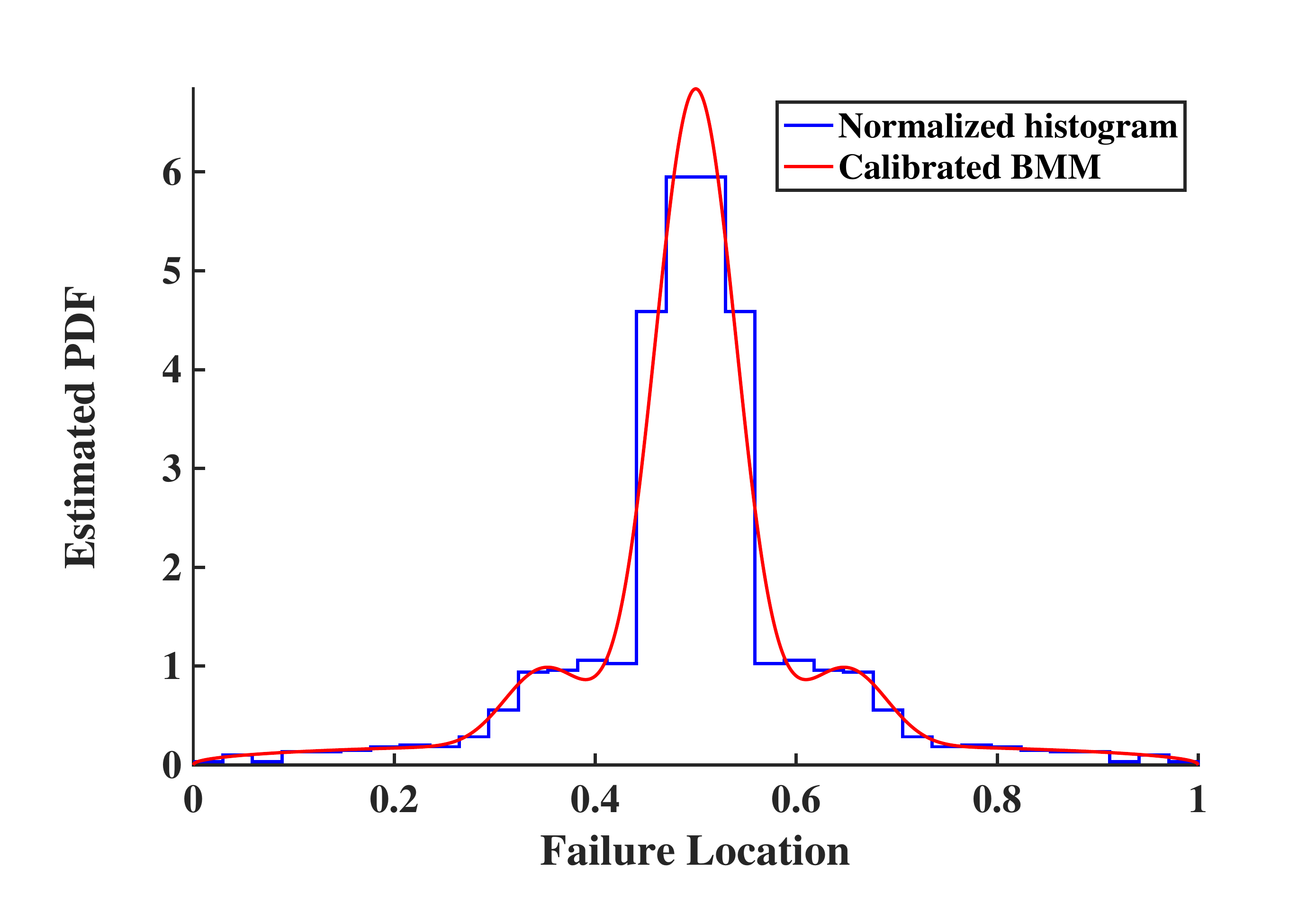}}
\qquad
\subfloat[Notch with 2.5 mm radius]
{\includegraphics[width=0.30\textwidth]{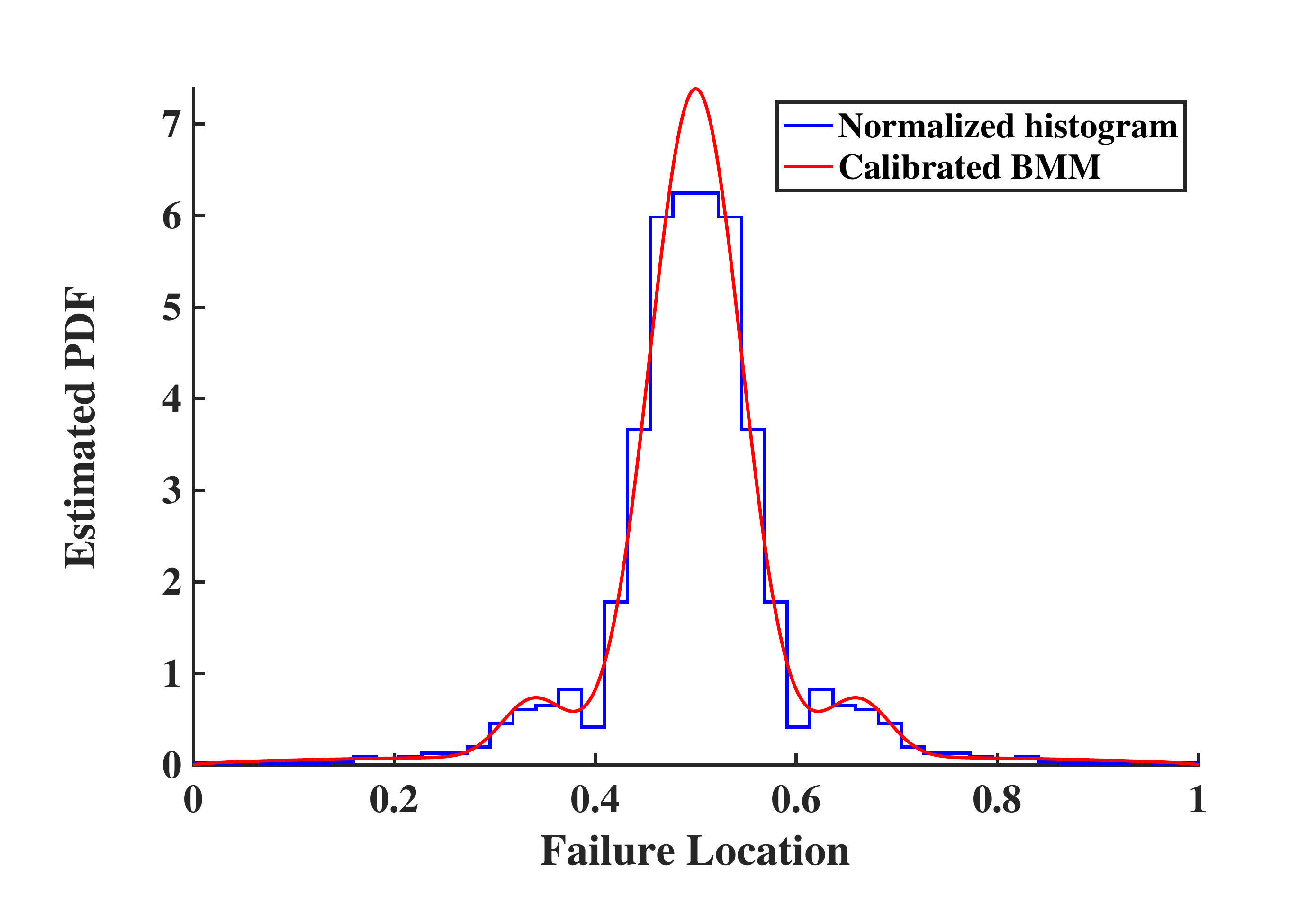}}

\subfloat[Notch with 3.75 mm radius]
{\includegraphics[width=0.30\textwidth]{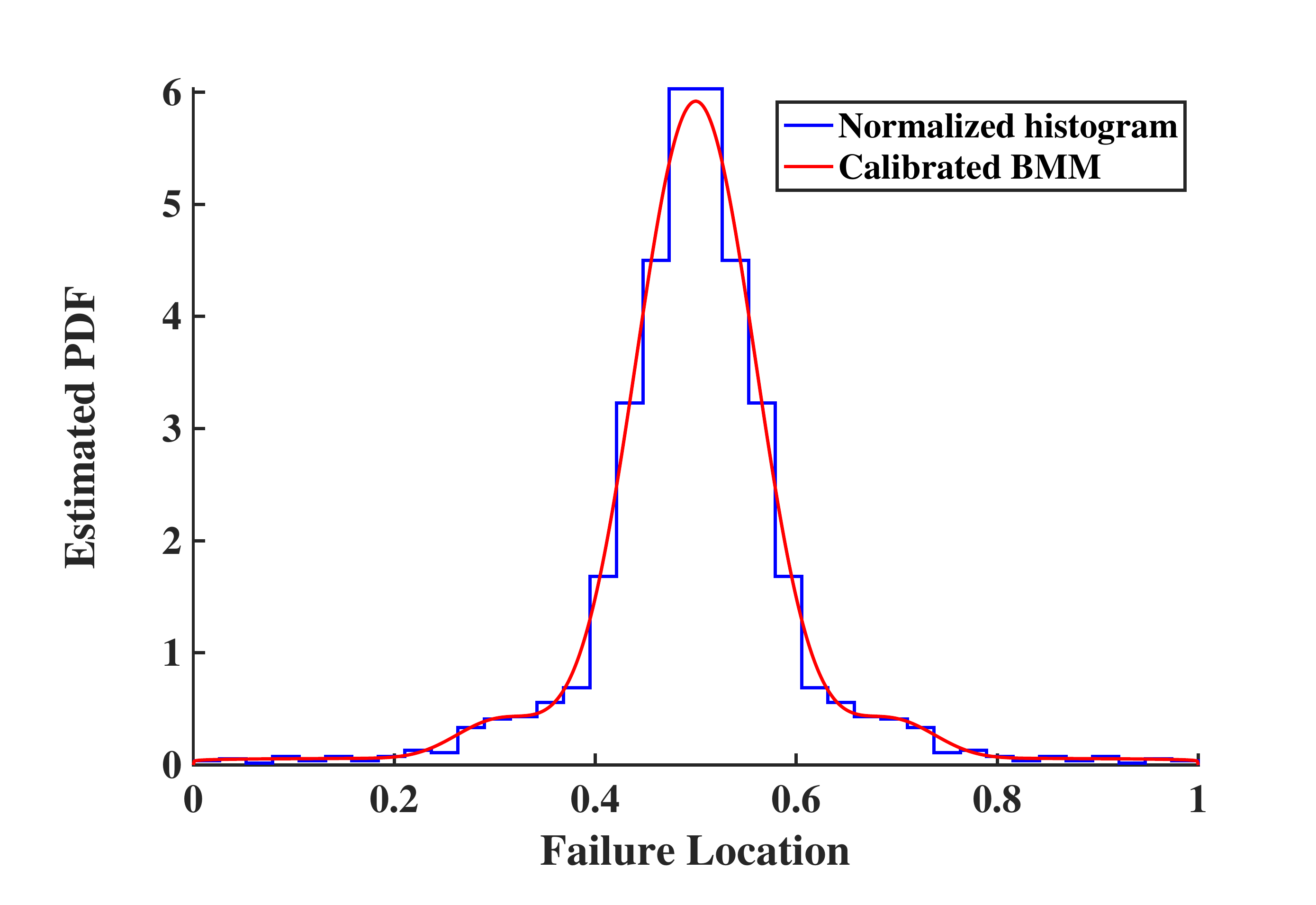}}
\qquad
\subfloat[Notch with 5 mm radius]
{\includegraphics[width=0.27\textwidth]{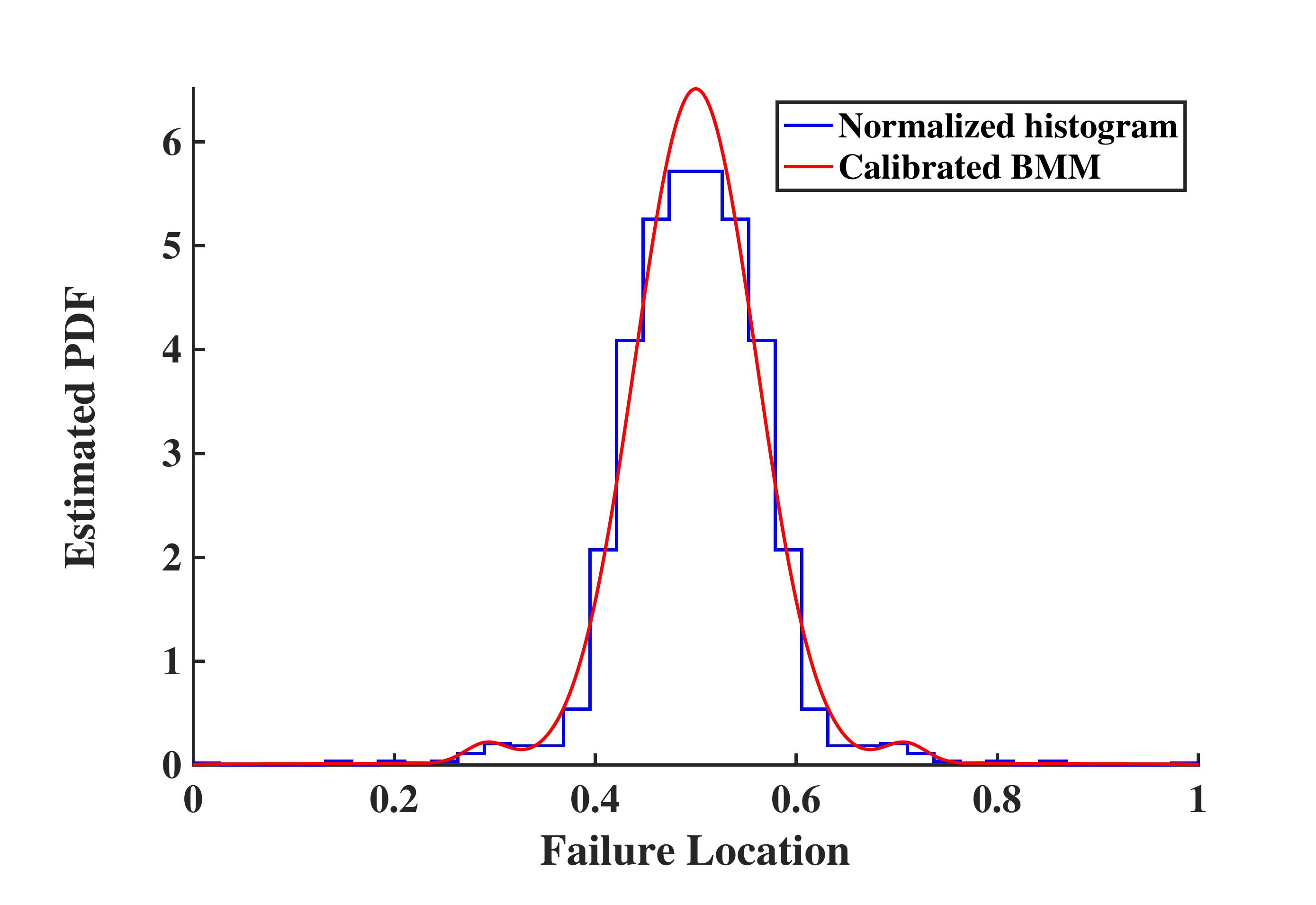}}
\qquad
\subfloat[Notch with 7.5 mm radius]
{\includegraphics[width=0.30\textwidth]{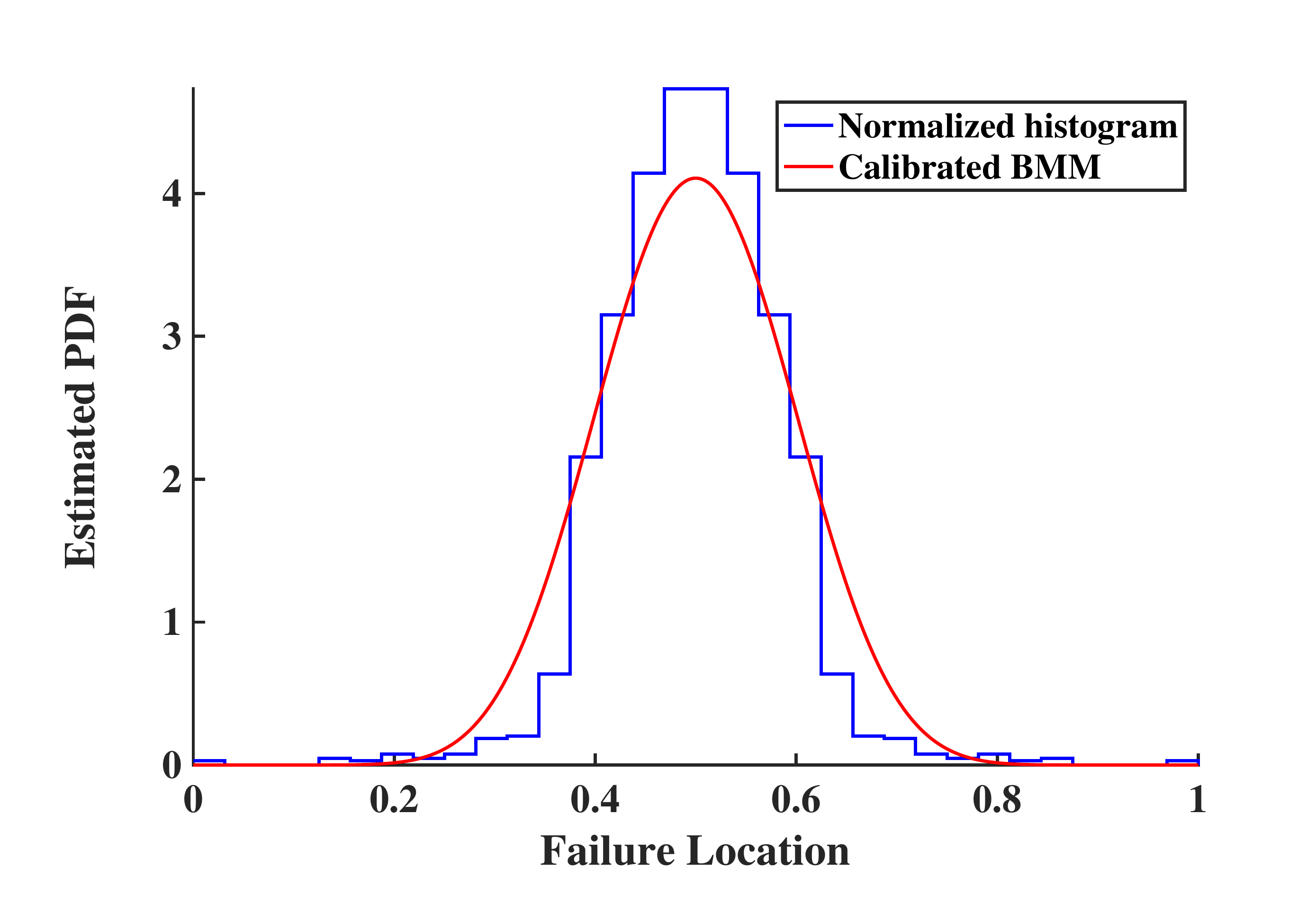}}

\subfloat[Notch with 10 mm radius]
{\includegraphics[width=0.30\textwidth]{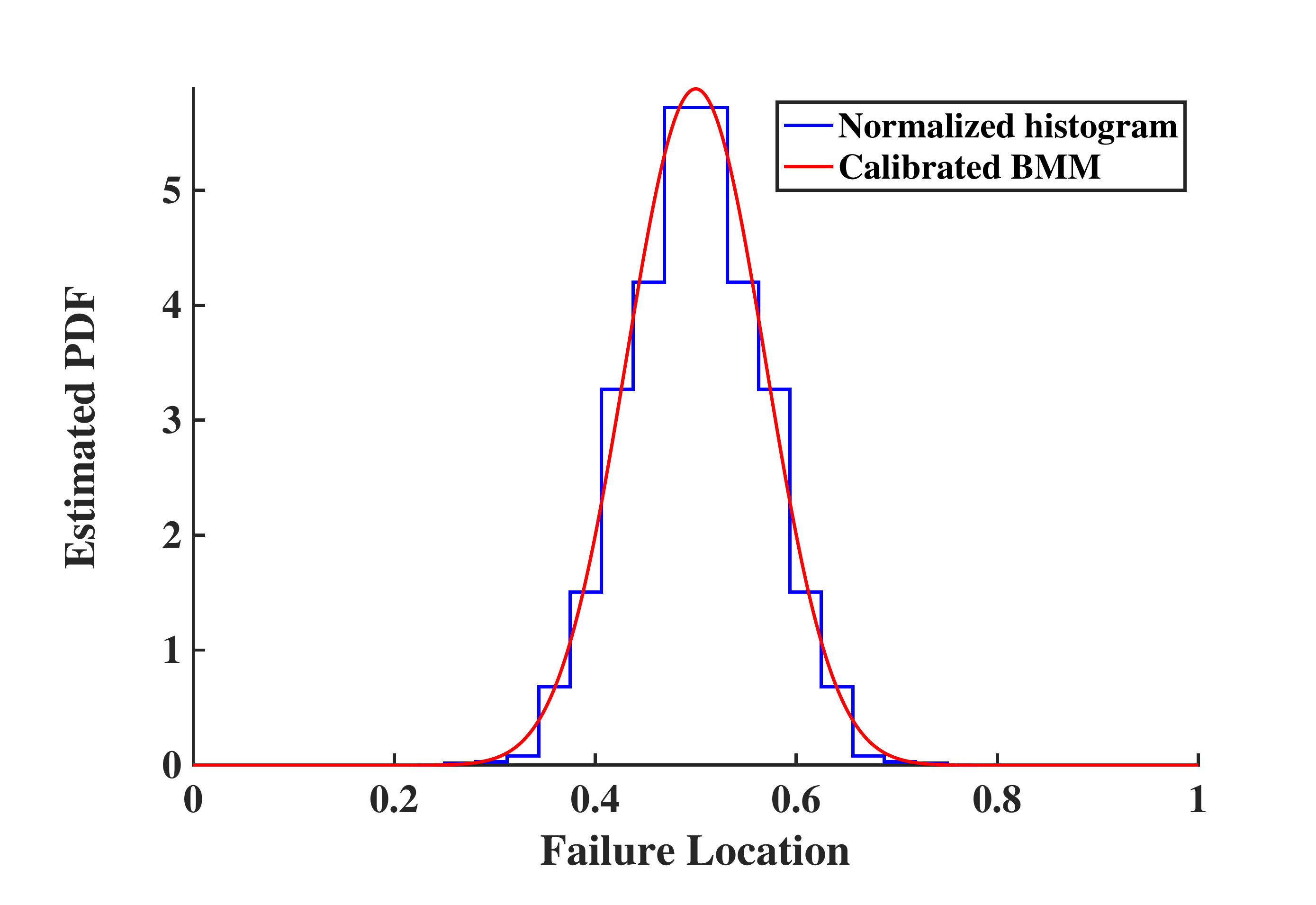}}
\qquad
\subfloat[Notch with 12.5 mm radius]
{\includegraphics[width=0.30\textwidth]{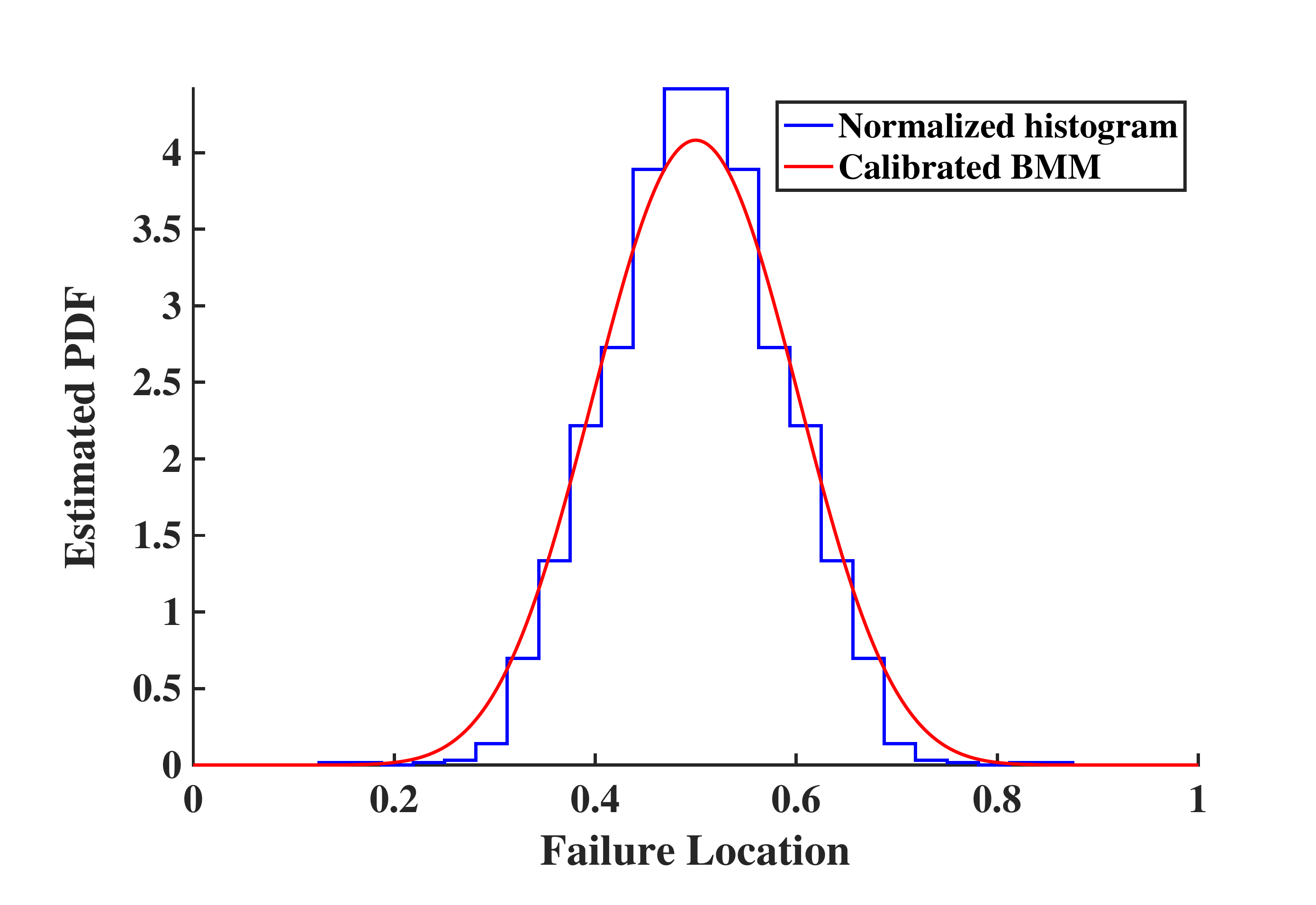}}
\qquad
\subfloat[Notch with 15 mm radius]
{\includegraphics[width=0.30\textwidth]{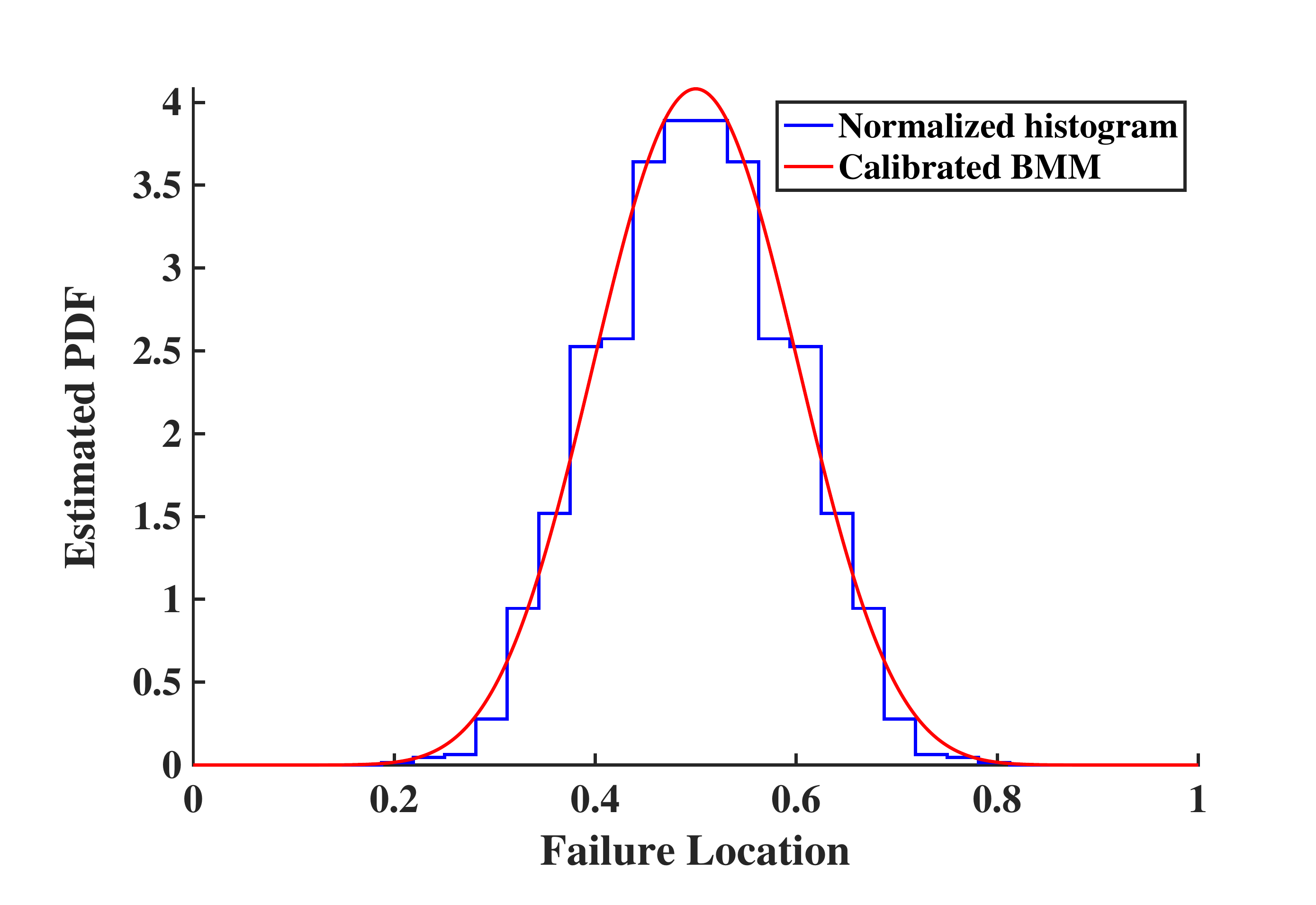}}

\caption{Fitted BMMs for various geometries with varying notch radius.
}
\label{fig:fitted_BMM}
\end{figure}

\fref{fig:fitted_BMM} shows the fitted mixture model superimposed over the histogram of failure locations for various notch radii.
The four component BMM represents the transition from the predominance of failure for sharply notched specimens to the nearly uniform expectation of failure for the no notch case.
Furthermore, both the broad, background propensity for failure due to the nearly-uniform distribution of pores/voids and secondary side modes that appear for intermediate notch radii are well captured.
For illustration, the components of the calibrated BMM, scaled by the component coefficients/weights, are shown in \fref{fig:fitted_BMM_components} for the notch of radius $R =$ 0.75 mm.
In what follows, we use the the term ``broad'' to refer to the first symmetric component; this component is sufficient to model the no notch case, and is thus associated with the background density. 
We use the term ``narrow'' to refer to the second symmetric component, with parameters $a=b$ of greater magnitude than the broad component; this component is associated with the high-triaxiality stress concentration aligned with the center/deepest part of the notch, as seen in \fref{fig:driving_fields:pressure}. 
We use the term ``asymmetric'' to refer to the mirrored off-center components associated with the shear-dominant/inclined stress concentration bands seen in \fref{fig:driving_fields:vonmises} emanating from the notches. 
The competitive action of these three drivers determines the evolution of the damage fields shown in \fref{fig:driving_fields:damage}. 
We quantify this competition in terms of the components of the BMM in the following section.



\clearpage

\subsection{Physical interpretation} \label{sec:physical}
With the mixture model we can answer the engineering design question, ``What is the likelihood of failure outside the notch region as a function of the stress intensities induced by the geometric features?"
The four-component mixture model was selected with the features of the empirical distributions shown in \fref{fig:hist_notch} in mind, namely: (a) a broad, nearly uniform distribution, (b) a narrow, peaked distribution centered at the notch, and (c) two minor, mirrored off-center peaks near the notch edges.
These component distributions can be easily interpreted as representative of three competing physical mechanisms: (a) statistically homogeneous stress concentrations arising from the material porosity, (b) a high-triaxiality stress concentration at the center of the notch driving a triaxial-nucleation$\to$growth$\to$coalescence failure process, and (c) shear-dominant stress concentrations emanating from the notch edges driving a shear-nucleation$\to$coalescence failure process.

\begin{figure}
\centering
\subfloat[Weight of the broad, background component \vs notch radius]
{\includegraphics[width=0.40\textwidth]{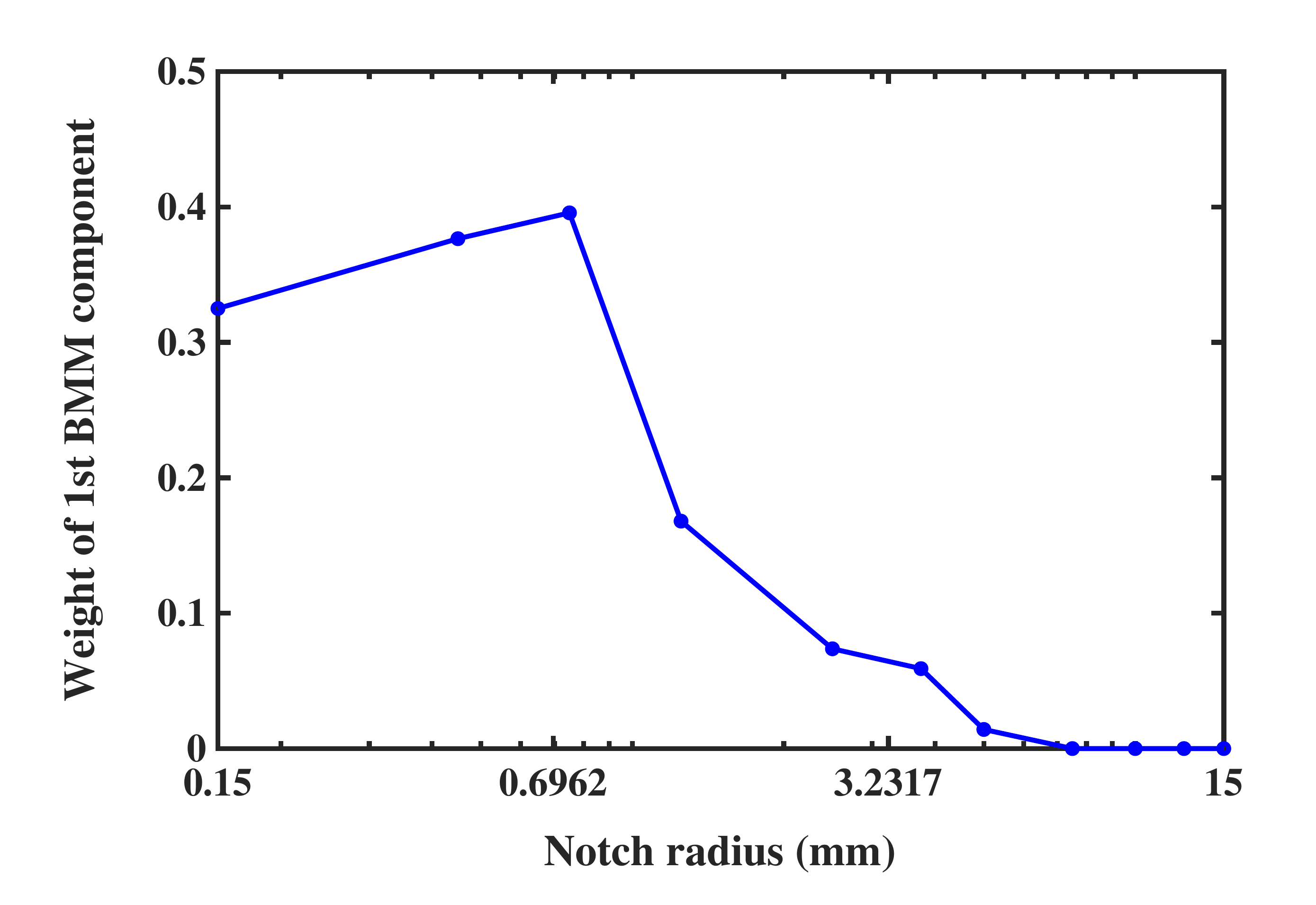}}
\qquad
\subfloat[Weight of the broad, background component \vs stress concentration]
{\includegraphics[width=0.40\textwidth]{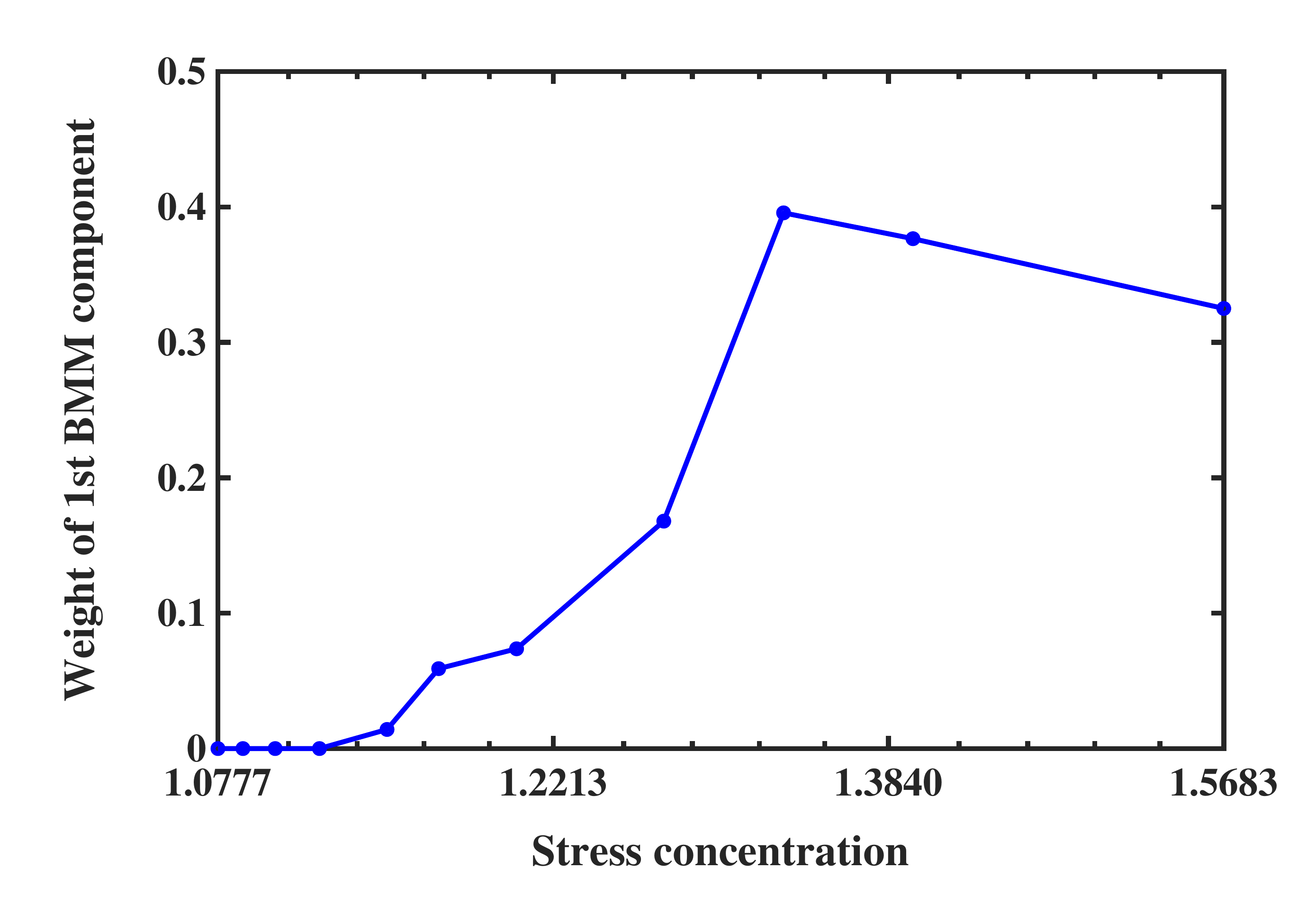}}

\subfloat[Weight of the narrow, centered component \vs notch radius]
{\includegraphics[width=0.40\textwidth]{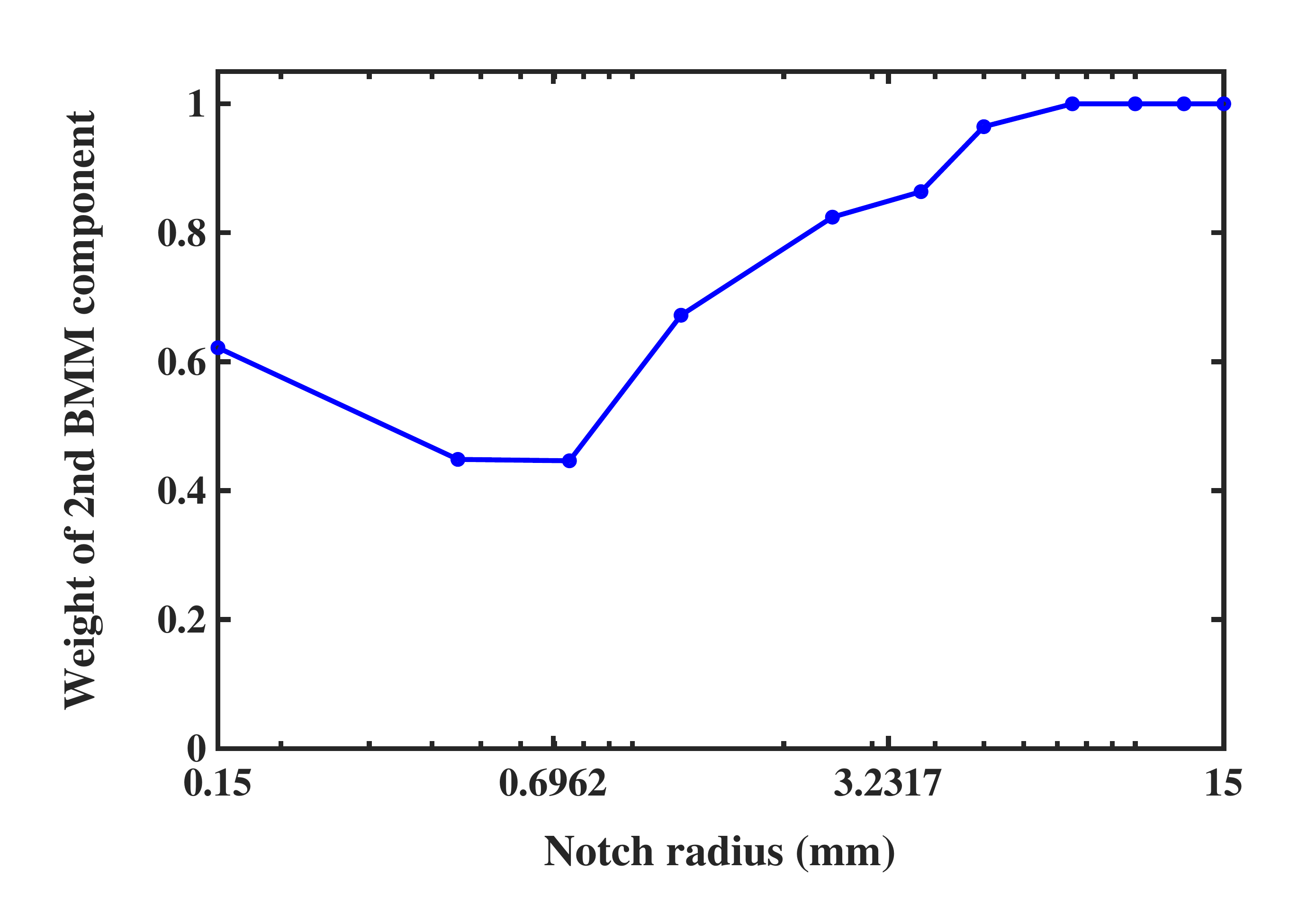}}
\qquad
\subfloat[Weight of the narrow, centered component \vs stress concentration]
{\includegraphics[width=0.40\textwidth]{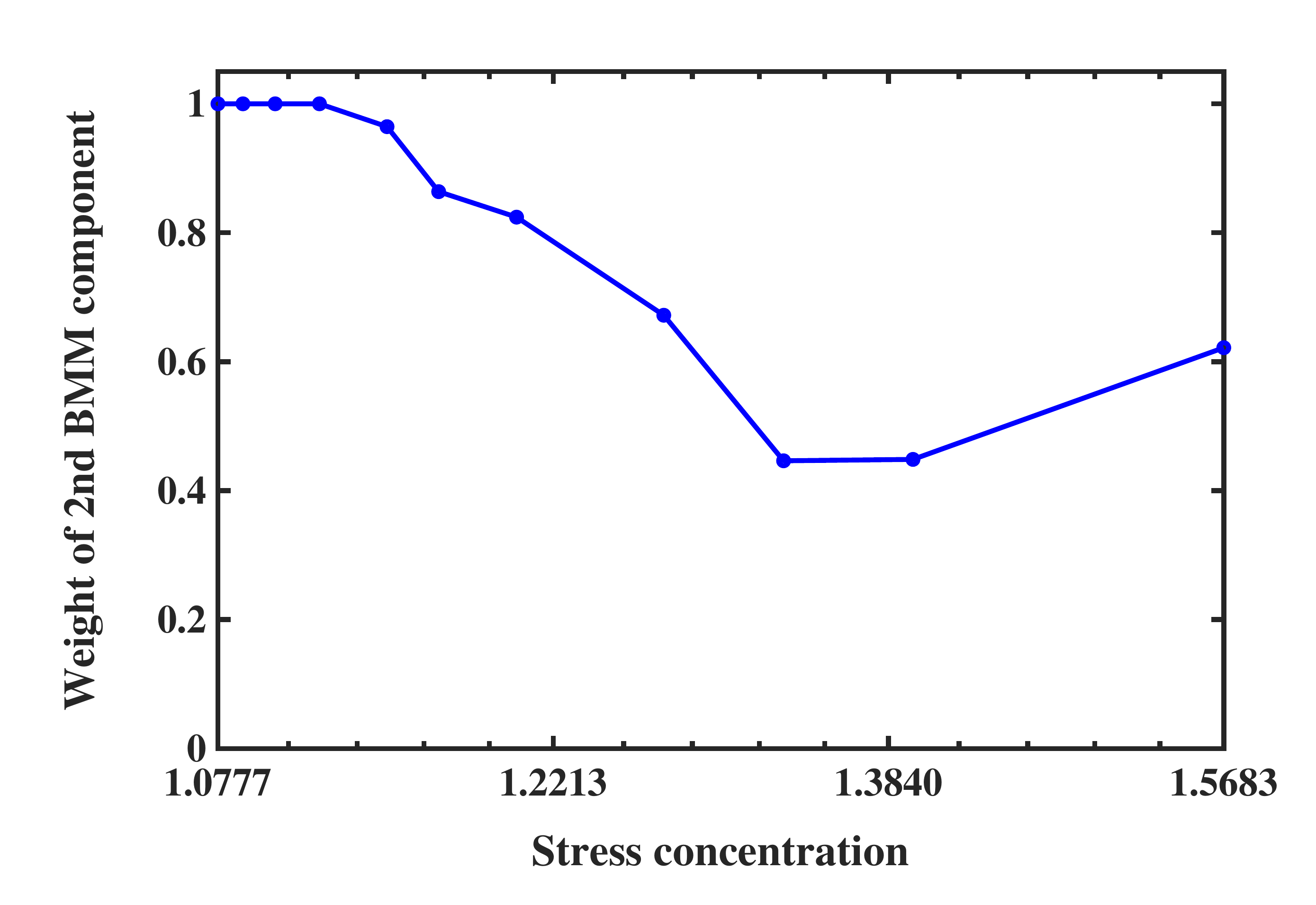}}

\subfloat[Weight of asymmetric, off-center component \vs notch radius]
{\includegraphics[width=0.40\textwidth]{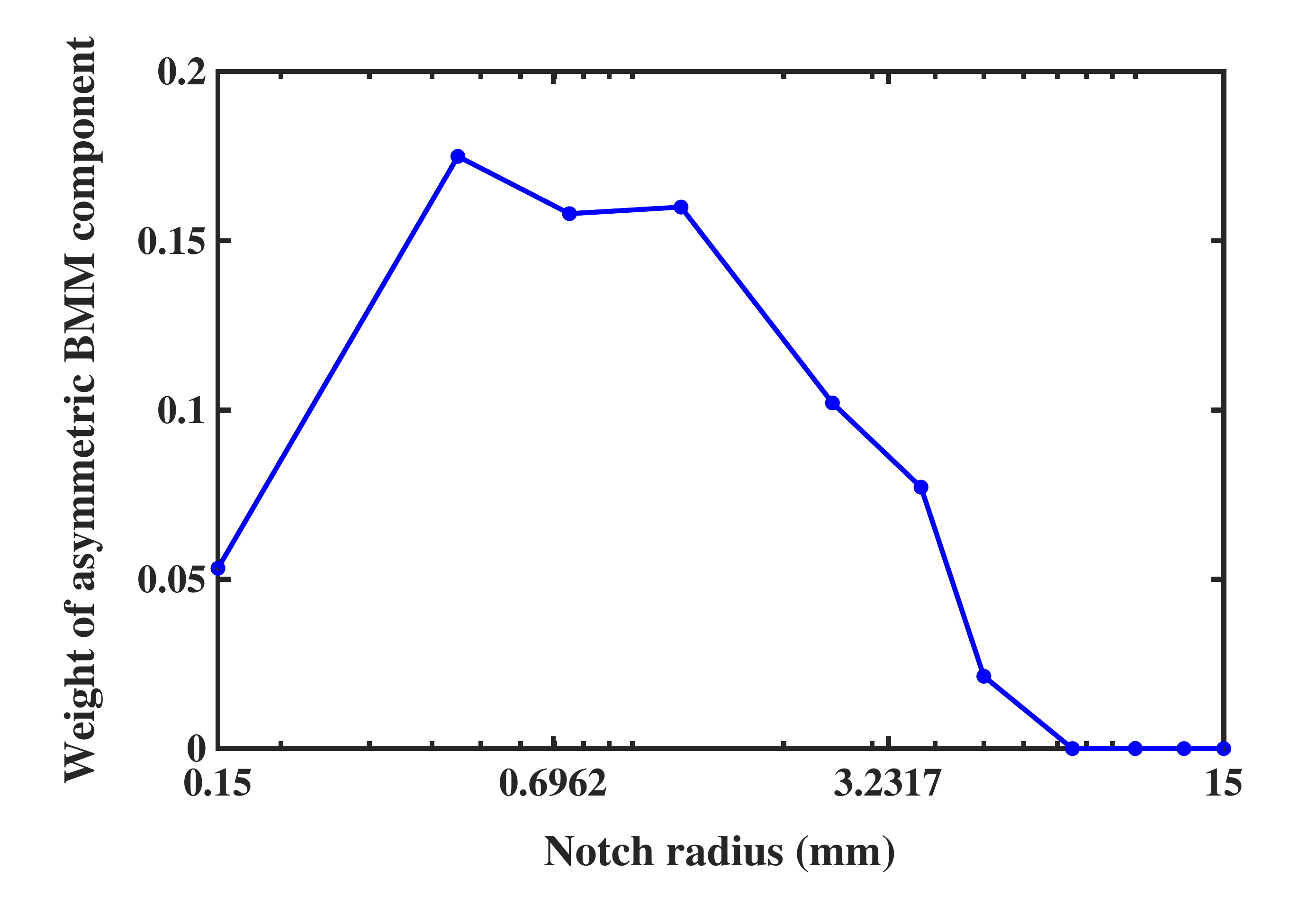}}
\qquad
\subfloat[Weight of the asymmetric, off-center component \vs stress concentration]
{\includegraphics[width=0.40\textwidth]{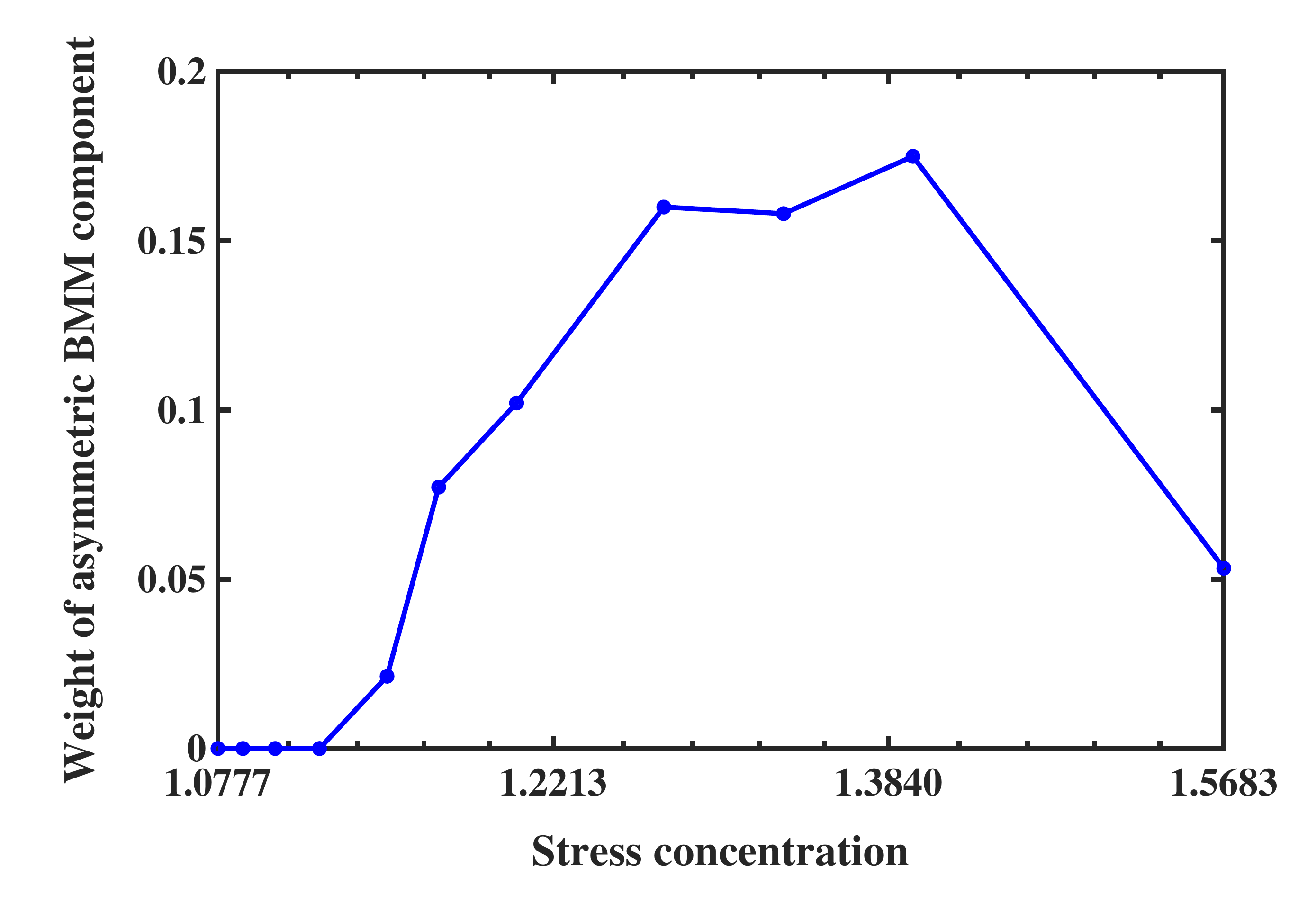}}

\caption{Weights of BMM components: (a-b) broad component, being symmetric, is intended to capture background scatter in failure location due to explicit porosity, (c-d) peaked component, also symmetric, captures failure due to notch and (e-f) asymmetric/off-center components capturing local cluster of failures around notch edge.
The weight reported for (e-f) is the combined weight of the two mirrored asymmetric components.
Note the notch radius and stress concentration are plotted on a log scale.
}
\label{fig:BMM_weights}
\end{figure}

\fref{fig:BMM_weights} shows the weights associated with the three mixture components, representing the relative importance of the three mechanisms of stress concentration in the failure processes for the different specimen geometries.
The weights are plotted as functions of both notch radius and stress concentration, which is inversely proportional to the notch radius per the empirical relation shown in \fref{fig:stress_concentration}.
A number of trends are evident in the plots in this figure; the conclusions are consistent whether the results are cast in terms of specimen geometry or in terms of stress concentration.

First, as the notch broadens the narrow component dominates while the broad and asymmetric/off-center components become negligible as the stress concentrations at the notch edges fade.
At the same time, the volume of the notch region grows which involves sufficient material to ensure that the failure triggering pore/s will be in this higher stress region.
In effect, there is increasing certainty that the specimen will fail within the notch as it becomes a taper; although some uncertainty in the precise location persists due to the stochastic nature of the explicitly represented porosity.

Second, for increasingly sharp notches there is a trend for the narrow component to dominate, but with significant contributions from the asymmetric component. 
Although not as well resolved due to limitations imposed by the mesh size we selected at the outset, this trend is clear for the sharper notches.
The stress concentrations at the notch edges are significant drivers for all but the sharpest notch, where the stress concentrations at the notch center becomes entirely dominant.

Third, there appears to be a transition regime where the notch is small enough relative to the specimen that failures occur with significant frequency outside the notch. 
This regime is where $R/L$ is on the order of $1/10$--$1$ for the notch depth of $D = 0.05$ mm analyzed in this section, which translates to stress intensities $\approx$ 1.2--1.5.
In much of this regime the off-center component is also significant, indicating that the effects of the secondary/shear stress concentrations at the notch edges are of comparable influence to the effects the material porosity.
From a design perspective \fref{fig:BMM_weights}d indicates the likelihood of failure outside the expected region as a function of stress concentration.
Clearly this can be as high as 40\%, but the effect is confined to a finite part of the design space as parameterized by the notch depth and radius.

These conclusions are based on a first order, additively decomposed BMM model where failures are categorized as either due to the notches' stress concentrations or the background of explicit pores.
A more sophisticated model could account for correlated effects; this development is left for future work.
The calibrated PDFs themselves have great utility, for instance they can be be integrated over any region of interest to estimate a probability of failure in that region.
Our technique could be used to analyze more general geometries, especially those where the dominant stress concentrating features are not known \apriori.
The continuity of the trends observed in the components of the BMM and how their features track the geometric features supports our claim that the association of the components of the BMM with physical characteristics of the specimens is valid.
In particular the standard deviation of the component representing the primary peak becomes essentially constant for $R/L > 1.5$ in concert with our observations and energy dissipation argument of \sref{sec:notched}.
Since the main reason we did not apply this technique to the full three dimensional failure location was insufficient data to resolve a three dimensional PDF, we believe the application of the technique in a more general setting will be straightforward.

\clearpage

\section{Conclusion} \label{sec:conclusion}

We performed an extensive statistical study to investigate how a background, uniform porosity interacts with stress concentrations due to as-designed, nominal geometry in a ductile AM metal.
Although we focused on our study on a single specimen geometry, the U-notched tension specimen is a commonly employed geometry and we generalized the results using the classical stress concentration concept.
The trends in the spatial distribution of failure locations are complex, and exhibit two asymptotic regimes: (a) where the stress concentration is high but the high stress region is small relative to the sample, and (b) where the higher stress region is large but the stress differences are relatively small. 
In the first case, a significant fraction of the failures are outside the stress concentration region due to the intrinsic stress concentrations of the pores themselves.
In the second case, effectively all failures occur in the notch region and the distribution of failure locations appears insensitive to geometry for specimens with well-separated geometric and material length scales.
Between these two regimes is an interesting regime where multi-modal distributions arise associated with the primary stress concentration at the notch root as well as secondary stress concentrations at the notch edges.
An interpretable statistical model fitted to the data allowed us to make these observations quantitative.
We found that the distribution of failures can be accurately described with a mixture of four beta distributions: (a) a broad one associated with the uniform density of pores, (b) a narrow one due to the strongest stress concentration at the notch root, and (c) two off-center, less dominant distributions associated with shear stress concentration regions near edges of the notch.
This work clearly demonstrates, for a relevant specimen geometry, the existence and extent of regimes where the likelihood of unexpected failure due to latent material defects is significant, and suggests paths to generalize these results in ways that could be used to establish quantitative design guidelines.

\section*{Acknowledgments}
The authors would like to acknowledge insightful discussions with Jay Foulk (Sandia) and Brad Boyce (Sandia).
Sandia National Laboratories is a multimission laboratory managed and operated by National Technology and Engineering Solutions of Sandia, LLC., a wholly owned subsidiary of Honeywell International, Inc., for the U.S. Department of Energy's National Nuclear Security Administration under contract DE-NA-0003525.
The views expressed in the article do not necessarily represent the views of the U.S. Department of Energy or the United States Government.

\typeout{}
\bibliography{porosity}

\begin{thebibliography}{10}
\expandafter\ifx\csname url\endcsname\relax
  \def\url#1{\texttt{#1}}\fi
\expandafter\ifx\csname urlprefix\endcsname\relax\def\urlprefix{URL }\fi
\expandafter\ifx\csname href\endcsname\relax
  \def\href#1#2{#2} \def\path#1{#1}\fi

\bibitem{heckman2020automated}
N.~M. Heckman, T.~A. Ivanoff, A.~M. Roach, B.~H. Jared, D.~J. Tung, H.~J.
  Brown-Shaklee, T.~Huber, D.~J. Saiz, J.~R. Koepke, J.~M. Rodelas, et~al.,
  Automated high-throughput tensile testing reveals stochastic process
  parameter sensitivity, Materials Science and Engineering: A 772 (2020)
  138632.

\bibitem{garino2004mechanical}
T.~J. Garino, A.~M. Morales, B.~L. Boyce, The mechanical properties,
  dimensional tolerance and microstructural characterization of micro-molded
  ceramic and metal components, Microsystem technologies 10~(6-7) (2004)
  506--509.

\bibitem{roach2020size}
A.~M. Roach, B.~C. White, A.~Garland, B.~H. Jared, J.~D. Carroll, B.~L. Boyce,
  Size-dependent stochastic tensile properties in additively manufactured 316l
  stainless steel, Additive Manufacturing 32 (2020) 101090.

\bibitem{brackett2011topology}
D.~Brackett, I.~Ashcroft, R.~Hague, Topology optimization for additive
  manufacturing, in: Proceedings of the solid freeform fabrication symposium,
  Austin, TX, Vol.~1, 2011, pp. 348--362.

\bibitem{deng2017concurrent}
J.~Deng, W.~Chen, Concurrent topology optimization of multiscale structures
  with multiple porous materials under random field loading uncertainty,
  Structural and Multidisciplinary Optimization 56~(1) (2017) 1--19.

\bibitem{vanderesse2018measurement}
N.~Vanderesse, A.~Richter, N.~Nu{\~n}o, P.~Bocher, Measurement of deformation
  heterogeneities in additive manufactured lattice materials by digital image
  correlation: Strain maps analysis and reliability assessment, Journal of the
  mechanical behavior of biomedical materials 86 (2018) 397--408.

\bibitem{solberg2019fatigue}
K.~Solberg, S.~Guan, S.~M.~J. Razavi, T.~Welo, K.~C. Chan, F.~Berto, Fatigue of
  additively manufactured 316l stainless steel: The influence of porosity and
  surface roughness, Fatigue \& Fracture of Engineering Materials \& Structures
  42~(9) (2019) 2043--2052.

\bibitem{delrio2020shoulder}
F.~W. DelRio, B.~L. Boyce, J.~T. Benzing, L.~H. Friedman, R.~F. Cook, Shoulder
  fillet effects in strength distributions of microelectromechanical system
  components, Journal of Micromechanics and Microengineering 30~(12) (2020)
  125013.

\bibitem{khalil2019modeling}
M.~Khalil, G.~Teichert, C.~Alleman, N.~Heckman, R.~Jones, K.~Garikipati,
  B.~Boyce, Modeling strength and failure variability due to porosity in
  additively manufactured metals, Computer Methods in Applied Mechanics and
  Engineering 373 (2021) 113471.

\bibitem{salzbrenner2017high}
B.~C. Salzbrenner, J.~M. Rodelas, J.~D. Madison, B.~H. Jared, L.~P. Swiler,
  Y.-L. Shen, B.~L. Boyce, High-throughput stochastic tensile performance of
  additively manufactured stainless steel, Journal of Materials Processing
  Technology 241 (2017) 1--12.

\bibitem{trustrum1979estimating}
K.~Trustrum, A.~D.~S. Jayatilaka, On estimating the weibull modulus for a
  brittle material, Journal of Materials Science 14~(5) (1979) 1080--1084.

\bibitem{trustrum1983applicability}
K.~Trustrum, A.~D.~S. Jayatilaka, Applicability of weibull analysis for brittle
  materials, Journal of Materials Science 18~(9) (1983) 2765--2770.

\bibitem{bazant1991statistical}
Z.~P. Bazant, Y.~Xi, S.~G. Reid, Statistical size effect in quasi-brittle
  structures: I. is weibull theory applicable?, Journal of engineering
  Mechanics 117~(11) (1991) 2609--2622.

\bibitem{fok2001numerical}
S.~Fok, B.~Mitchell, J.~Smart, B.~Marsden, A numerical study on the application
  of the weibull theory to brittle materials, Engineering Fracture Mechanics
  68~(10) (2001) 1171--1179.

\bibitem{becker1988void}
R.~Becker, A.~Needleman, O.~Richmond, V.~Tvergaard, Void growth and failure in
  notched bars, Journal of the Mechanics and Physics of Solids 36~(3) (1988)
  317--351.

\bibitem{jadaan2003probabilistic}
O.~M. Jadaan, N.~N. Nemeth, J.~Bagdahn, W.~Sharpe, Probabilistic weibull
  behavior and mechanical properties of mems brittle materials, Journal of
  materials science 38~(20) (2003) 4087--4113.

\bibitem{mccarty2007description}
A.~McCarty, I.~Chasiotis, Description of brittle failure of non-uniform mems
  geometries, Thin solid films 515~(6) (2007) 3267--3276.

\bibitem{cook2019predicting}
R.~F. Cook, F.~W. DelRio, B.~L. Boyce, Predicting strength distributions of
  mems structures using flaw size and spatial density, Microsystems \&
  nanoengineering 5~(1) (2019) 1--12.

\bibitem{rice1993comparison}
R.~Rice, Comparison of stress concentration versus minimum solid area based
  mechanical property-porosity relations, Journal of materials science 28~(8)
  (1993) 2187--2190.

\bibitem{hardin2013effect}
R.~Hardin, C.~Beckermann, Effect of porosity on deformation, damage, and
  fracture of cast steel, Metallurgical and Materials Transactions A 44~(12)
  (2013) 5316--5332.

\bibitem{yin2009efficient}
X.~Yin, S.~Lee, W.~Chen, W.~K. Liu, M.~Horstemeyer, Efficient random field
  uncertainty propagation in design using multiscale analysis, Journal of
  Mechanical Design 131~(2) (2009).

\bibitem{allison2013plasticity}
P.~G. Allison, H.~Grewal, Y.~Hammi, H.~R. Brown, W.~R. Whittington, M.~F.
  Horstemeyer, Plasticity and fracture modeling/experimental study of a porous
  metal under various strain rates, temperatures, and stress states, Journal of
  engineering materials and technology 135~(4) (2013).

\bibitem{shakoor2018ductile}
M.~Shakoor, M.~Bernacki, P.-O. Bouchard, Ductile fracture of a metal matrix
  composite studied using 3d numerical modeling of void nucleation and
  coalescence, Engineering Fracture Mechanics 189 (2018) 110--132.

\bibitem{boyce2017extreme}
B.~L. Boyce, B.~C. Salzbrenner, J.~M. Rodelas, L.~P. Swiler, J.~D. Madison,
  B.~H. Jared, Y.-L. Shen, Extreme-value statistics reveal rare
  failure-critical defects in additive manufacturing, Advanced Engineering
  Materials 19~(8) (2017) 1700102.

\bibitem{bammann1996modeling}
D.~Bammann, M.~Chiesa, G.~Johnson, Modeling large deformation and failure in
  manufacturing processes, Theoretical and Applied Mechanics 9 (1996) 359--376.

\bibitem{horstemeyer1999void}
M.~Horstemeyer, A.~Gokhale, A void--crack nucleation model for ductile metals,
  International Journal of Solids and Structures 36~(33) (1999) 5029--5055.

\bibitem{brown2012validation}
A.~Brown, D.~Bammann, Validation of a model for static and dynamic
  recrystallization in metals, International Journal of Plasticity 32 (2012)
  17--35.

\bibitem{karlson2016sandia}
K.~Karlson, J.~Foulk, A.~Brown, M.~Veilleux, Sandia fracture challenge 2:
  {S}andia {C}alifornia’s modeling approach, International Journal of
  Fracture 198~(1-2) (2016) 179--195.

\bibitem{boyce2013morphology}
B.~L. Boyce, B.~G. Clark, P.~Lu, J.~D. Carroll, C.~R. Weinberger, The
  morphology of tensile failure in tantalum, Metallurgical and Materials
  Transactions A 44~(10) (2013) 4567--4580.

\bibitem{furnish2016fatigue}
T.~A. Furnish, B.~L. Boyce, J.~A. Sharon, C.~J. O’Brien, B.~G. Clark, C.~L.
  Arrington, J.~R. Pillars, Fatigue stress concentration and notch sensitivity
  in nanocrystalline metals, Journal of Materials Research 31 (2016).

\bibitem{pilkey1997petersons}
W.~Pilkey, Petersons Stress Concentration Factors, John Wiley \& Sons, New
  York, 1997.

\bibitem{schijve2001fatigue}
J.~Schijve, Fatigue of structures and materials, Springer Science \& Business
  Media, 2001.

\bibitem{everitt2014finite}
B.~S. Everitt, Finite mixture distributions, Wiley StatsRef: Statistics
  Reference Online (2014).

\bibitem{lindsay1995mixture}
B.~G. Lindsay, Mixture models: theory, geometry and applications, in: NSF-CBMS
  regional conference series in probability and statistics, JSTOR, 1995, pp.
  i--163.

\bibitem{marin2005bayesian}
J.-M. Marin, K.~Mengersen, C.~P. Robert, Bayesian modelling and inference on
  mixtures of distributions, Handbook of statistics 25 (2005) 459--507.

\bibitem{mclachlan2004finite}
G.~J. McLachlan, D.~Peel, Finite mixture models, John Wiley \& Sons, 2004.

\bibitem{titterington1985statistical}
D.~M. Titterington, A.~F. Smith, U.~E. Makov, Statistical analysis of finite
  mixture distributions, Wiley, 1985.

\bibitem{bouguila2006practical}
N.~Bouguila, D.~Ziou, E.~Monga, Practical bayesian estimation of a finite beta
  mixture through gibbs sampling and its applications, Statistics and Computing
  16~(2) (2006) 215--225.

\bibitem{ma2011bayesian}
Z.~Ma, A.~Leijon, Bayesian estimation of beta mixture models with variational
  inference, IEEE Transactions on Pattern Analysis and Machine Intelligence
  33~(11) (2011) 2160--2173.

\bibitem{eliason1993maximum}
S.~R. Eliason, Maximum likelihood estimation: Logic and practice, Sage, 1993.

\bibitem{fan1998local}
J.~Fan, M.~Farmen, I.~Gijbels, Local maximum likelihood estimation and
  inference, Journal of the Royal Statistical Society: Series B (Statistical
  Methodology) 60~(3) (1998) 591--608.

\bibitem{schoen1991stochastic}
F.~Schoen, Stochastic techniques for global optimization: A survey of recent
  advances, Journal of Global Optimization 1~(3) (1991) 207--228.

\bibitem{lagarias1998convergence}
J.~C. Lagarias, J.~A. Reeds, M.~H. Wright, P.~E. Wright, Convergence properties
  of the nelder--mead simplex method in low dimensions, SIAM Journal on
  optimization 9~(1) (1998) 112--147.

\bibitem{akaike1998information}
H.~Akaike, Information theory and an extension of the maximum likelihood
  principle, in: Selected papers of hirotugu akaike, Springer, 1998, pp.
  199--213.

\bibitem{ghosh2007introduction}
J.~K. Ghosh, M.~Delampady, T.~Samanta, An introduction to Bayesian analysis:
  theory and methods, Springer Science \& Business Media, 2007.

\bibitem{roeder1997practical}
K.~Roeder, L.~Wasserman, Practical bayesian density estimation using mixtures
  of normals, Journal of the American Statistical Association 92~(439) (1997)
  894--902.

\bibitem{corduneanu2001variational}
A.~Corduneanu, C.~M. Bishop, Variational bayesian model selection for mixture
  distributions, in: Artificial intelligence and Statistics, Vol. 2001, Morgan
  Kaufmann Waltham, MA, 2001, pp. 27--34.

\bibitem{gilpin2018explaining}
L.~H. Gilpin, D.~Bau, B.~Z. Yuan, A.~Bajwa, M.~Specter, L.~Kagal, Explaining
  explanations: An overview of interpretability of machine learning, in: 2018
  IEEE 5th International Conference on data science and advanced analytics
  (DSAA), IEEE, 2018, pp. 80--89.

\end{thebibliography}

\end{document}